\documentclass[twocolumn,journal,10pt]{IEEEtran}
\usepackage{latexsym, amssymb, amsfonts,amsmath,mathrsfs,amstext,graphicx, color, bm, times, hyperref, relsize, mathptm, soul,amsthm,float}

\usepackage{algorithm}
\usepackage{algpseudocode}

\floatname{algorithm}{Protocol}
 \def\map#1{\mathscr #1}
\def\Choi{\operatorname{Choi}}
\def\d{\operatorname{d}}\def\<{\langle}\def\>{\rangle}
\def\bb{\langle\!\langle}\def\kk{\rangle\!\rangle}
\def\Tr{\operatorname{Tr}}\def\:{\hbox{\bf :}}

\def\vec#1{\mathbf{#1}}
\def\R{\mathbb R}
\def\N{\mathbb N}

\def\Span{\mathsf{Span}}

\def\spc#1{\mathcal{#1}}

\def\set#1{\mathcal{#1}}

\def\Proof{{\bf Proof.}}

\newtheorem{theo}{{Theorem}}
\newtheorem{lem}{{Lemma}}
\newtheorem{prop}{{Proposition}}
\newtheorem{cor}{{Corollary}}
\newtheorem{cond}{{Condition}}

\newcommand{\red}[1]{{{#1}}}

\begin{document}
\title{Communication cost of quantum processes}
\author{Yuxiang Yang, Giulio Chiribella,  and~Masahito~Hayashi,~\IEEEmembership{Fellow,~IEEE}
\thanks{This work is supported by the National Natural Science Foundation of China through Grant No. 11675136, by the Swiss National Science Foundation via the National Center for Competence in Research ``QSIT" as well as via project No.\ 200020\_165843, by the AFOSR via grant No. FA9550-19-1-0202, by the Hong Kong Research Grant Council
through Grant Nos. 17300317 and 17300918, and
by the HKU Seed Funding for Basic Research, and 
by the Foundational Questions Institute through grant FQXi-RFP3-1325. 
MH was supported in part by 
Guangdong Provincial Key Laboratory (Grant No. 2019B121203002),
a JSPS Grant-in-Aid for Scientific Research 
(A) No.17H01280, (B) No. 16KT0017, the Okawa Research Grant and Kayamori
Foundation of Informational Science Advancement. 
YY gratefully acknowledges the hospitality of South University of Science and Technology of China during the completion of part of this work. 
Research at Perimeter Institute is supported by the Government of Canada through the Department of Innovation, Science and Economic Development Canada and by the Province of Ontario through the Ministry of Research, Innovation and Science.}

\thanks{Yuxiang Yang is with Institute for Theoretical Physics, ETH Z\"urich.
(e-mail: yangyu@phys.ethz.ch)}
\thanks{Giulio Chiribella is with
QICI Quantum Information and Computation Initiative, Department of Computer Science, The University of Hong Kong, Pokfulam Road, Hong Kong,
Department of Computer Science, University of Oxford, Parks Road, Oxford, UK,
HKU Shenzhen Institute of Research and Innovation, Kejizhong 2nd Road, Shenzhen, 518057, China, and
Perimeter Institute For Theoretical Physics, 31 Caroline Street North,  Waterloo N2L 2Y5, Ontario, Canada.
(e-mail: giulio@cs.hku.hk)}
\thanks{Masahito Hayashi is with
Shenzhen Institute for Quantum Science and Engineering, Southern University of Science and Technology,
Shenzhen, 518055, China,
Guangdong Provincial Key Laboratory of Quantum Science and Engineering,
Southern University of Science and Technology, Shenzhen 518055, China,
Shenzhen Key Laboratory of Quantum Science and Engineering, Southern
University of Science and Technology, Shenzhen 518055, China,
and
the Graduate School of Mathematics, Nagoya University, Nagoya, 464-8602, Japan
(e-mail: hayashi@sustech.edu.cn)}}

\markboth{Y. Yang, G. Chiribella,  and~M.~Hayashi: Communication cost of quantum processes}{}

\maketitle

\begin{abstract}
A common scenario in distributed computing involves a client who asks a server to perform a  computation on a remote computer.   An important problem is   to determine  the minimum amount of communication  needed to specify the desired computation.   Here we extend this problem to the quantum domain, analyzing  the total amount of (classical and quantum) communication needed by a server in order to accurately execute a  quantum process chosen by a client from a parametric family of quantum processes.   We derive a general lower bound on the communication cost,  establishing a relation with the precision limits of quantum metrology:  if a $v$-dimensional family of processes can be estimated with  mean squared error $n^{-\beta}$ by using $n$ parallel queries, then the communication cost for $n$ parallel executions of a process in the family is  at least $(\beta \, v /2-\epsilon) \, \log n$ qubits at the leading order in $n$, for every  $\epsilon>0$.  For a class of quantum processes satisfying the standard quantum limit ($\beta  = 1$), we show that the bound can be attained by transmitting an approximate classical description of the desired process.   For quantum processes satisfying the Heisenberg limit ($\beta=2$), our bound shows that the communication cost is at least twice as the cost of communicating standard quantum limited processes with the same number of parameters. 
\end{abstract}


\section{Introduction}
Quantifying the  communication cost for the  execution of a desired computation on a remote computer is a fundamental issue in classical distributed computing \cite{Yao1979}. It informs the design of distributed algorithms \cite{lynch1996distributed} and wireless sensor networks \cite{dargie2010fundamentals}. 
Similar issues arise also  in quantum computing, in particular in the tasks of quantum gate teleportation \cite{gottesman1999demonstrating,PhysRevA.58.2745}   and delegated  quantum computation \cite{childs2005secure,broadbent2009universal,fitzsimons2017unconditionally,kashefi2017multiparty}, when one party is asked  to apply a sequence of quantum gates on an input state remotely provided  by another party. 
More generally, determining the amount of communication needed to specify a desired quantum process  is relevant to a number of information-theoretic tasks, including the design of programmable quantum devices \cite{nielsen1997programmable,  fiuravsek2002universal, bergou2005universal, hillery2006approximate, d2005efficient,ishizaka2008asymptotic,sedlak2019optimal,kubicki2019resource},  the  conversion of quantum gates \cite{chiribella2008optimal,dur2015deterministic,gate-replication,chiribella2016optimal,miyazaki2017universal,yang2017units}, quantum process tomography \cite{chuang1997prescription,poyatos1997complete,altepeter2003ancilla}, and quantum reading \cite{pirandola2011quantum,pirandola2011quantum2}.


In this paper we analyze the scenario where  a client uses a quantum communication link to specify a quantum process chosen from a given parametric family.   The client sends out a  (generally quantum) message that provides the server with a description of the desired process. The server then uses the client's message as a program to approximately implement the desired process for $n\ge 1$ times in parallel on $n$ identical systems. 
The  approximation error, the number of applications $n$, and the parametric family containing the desired process are assumed to be known both to the client and the server, who  use this knowledge to minimize the amount of communication needed to fulfil their task.

Our paper contains two main results. The first main result is a lower bound that relates the communication cost  with the precision limits of quantum metrology \cite{holevo,helstrom,giovannetti2006quantum,giovannetti2011advances}.  Precisely, we show that, if a   family of processes  can be estimated with mean squared error $n^{-\beta}$, then such family has  a communication cost of at least $(\beta \,  v  /2-\epsilon) \, \log n$ qubits at the leading order in $n$, where $v$ is   the number of   real parameters parametrizing the processes in the given family and $\epsilon$ is an arbitrary positive number. Our lower  bound also applies to a more general setting where the client executes a compressed version of the original process, and the server retrieves the original process  by applying suitable pre- and post-processing operations.        

The second main result is an achievability result. For a class of quantum processes satisfying the standard quantum limit (mean squared error vanishing as $1/n$), we show that our lower bound can be attained by transmitting an approximate classical description of the desired process.   For quantum processes satisfying the Heisenberg limit (mean squared error vanishing as $1/n^2$ \cite{luis1996optimum,buvzek1999optimal,chiribella2004efficient,bagan2004quantum,hayashi2006parallel,giovannetti2006quantum}), our lower  bound shows that the communication cost is at least twice as the cost of communicating standard quantum limited processes with the same number of parameters.



\section{Notation}

For a Hilbert space $\spc H$ and a vector $|\psi\>  \in  \spc H$, we will use the notation $\psi:  =  |\psi\>\<\psi|$ to denote the projector on the one-dimensional subspace spanned by $|\psi\>$.  

The space of linear operators from a Hilbert space $\spc H$ to another  Hilbert space $\spc K$  will be denoted by $L(\spc H,  \spc K)$.  When the two Hilbert spaces coincide, we will use the shorthand  $L(\spc H): =  L(\spc H,\spc H)$.  In this paper we will focus on finite-dimensional quantum systems,  with $\dim  (\spc H)  <\infty$.  

For a quantum system with Hilbert space $\spc H$, the set of quantum states   will be denoted by  $\operatorname{St} (\spc H) :  = \{  \rho  \in  L(\spc H) ~|~  \Tr [\rho]=1 \, , \,    \<\psi|  \rho  |\psi\>  \ge 0  \,  \forall |\psi\> \in \spc H \}$.    The subset of pure states (rank-one projectors) will be denoted as $\operatorname{PurSt} (\spc H) :  =  \{  |\psi\>\<\psi|  ~|~ |\psi\>  \in  \spc H  \,,   \<\psi|  \psi\> =1\}$.

 A quantum process transforming an input system into a (possibly different) output system is called a {\em quantum channel}.   A quantum channel transforming an  input system with Hilbert space $\spc H_{\rm in}$ into an output system with (possibly different) Hilbert space $\spc H_{\rm out}$ is a completely positive trace-preserving map $\map C:  L( \spc H_{\rm in})  \to L  (\spc H_{\rm out})$.  The set of all quantum channels with input space $\spc H_{\rm in}$ and output space $\spc H_{\rm out}$ will be denoted by  ${\sf Chan}  (\spc H_{\rm in},\spc H_{\rm out})$.

 We stress  that ``quantum channel'' is a technical term,  and it does not generally refer to a quantum {\em communication}   channel placed between a sender and a receiver.   In this paper, all the quantum communication channels between client and server will be assumed to be noiseless and will be left implicit. All the quantum channels explicitly described in the paper  represent  quantum processes happening either at the client's or at server's end. 
  
The space of linear maps from $L( \spc H_{\rm in})$ to $L  (\spc H_{\rm out})$  will be denoted as ${\sf Map} (\spc H_{\rm in}, \spc H_{\rm out})  :  =  L  \big( L( \spc H_{\rm in}) \, ,  L  (\spc H_{\rm out})  \big) $, again with the convention ${\sf Map}  (\spc H)  :  =  {\sf Map}  (\spc H, \spc H)$.     

The space of linear maps   ${\sf Map} (\spc H_{\rm in}, \spc H_{\rm out}) $   is in one-to-one correspondence with the space of linear operators $L(  \spc H_{\rm out}\otimes\spc H_{\rm in})$ via the Choi correspondence   
  \cite{choi1975completely}  
\begin{align}
{\sf Choi}:&     {\sf Map} (\spc H_{\rm in}, \spc H_{\rm out})     \to   L(  \spc H_{\rm out}\otimes\spc H_{\rm in})  \nonumber \\
  & \map M    \mapsto  \Choi ({\map{M}}):=(\map{M}\otimes\map{I}_{{\rm in}})(|I\kk\bb I|)
\end{align}
where $\map I_{\rm in}  \in {\sf Map } (\spc H_{\rm in})$   is the identity map,  and  $|I\kk   \in \spc H_{\rm in} \otimes \spc H_{\rm in}$ is the unnormalized maximally entangled state $| I \kk  :=\sum_{k}|k\>\otimes |k\>$, defined in terms of a fixed (but otherwise arbitrary) orthonormal basis  $\{|k\>\}$ for $\spc{H}_{\rm in}$.

For a quantum channel $\map C  \in  {\sf Chan}  (\spc H_{\rm in}, \spc H_{\rm out})$, the Choi operator satisfies the normalization condition 
\begin{align}\label{choinormalization}
\Tr_{\rm out}    [   \Choi  (\map C)  ]    =  I_{\rm in} \, ,
\end{align}
where $\Tr_{\rm out}$ denotes the partial trace over $\spc  H_{\rm out}$ and $I_{\rm in}$ denotes the identity operator on $\spc H_{\rm in}$. 

\section{The channel communication task}\label{sec:chancommun} 

\begin{figure}  [t!]
\begin{center}
  \includegraphics[width=0.95\linewidth]{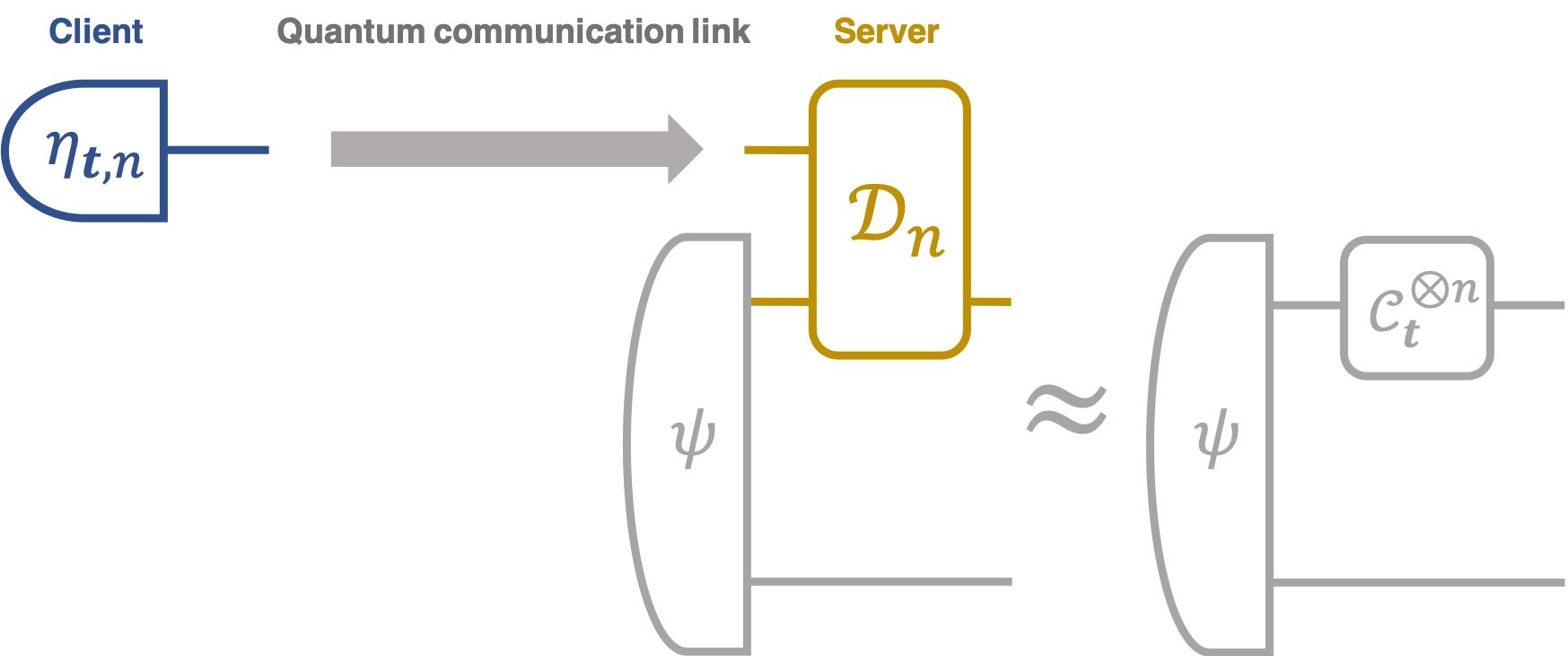}
  \end{center}
\caption{\label{fig-visible}
  {\bf The channel communication task.}  A client requires a server to perform a quantum channel  $\map{C}_{\vec{t}}$ for $n$ times in parallel.   To this purpose, the client   sends a program $\eta_{\vec{t},n}$ (in blue) to the server. Then, the server decodes the program by performing a decoding channel $\map{D}_n$ (in gold), which approximates the target channel $\map{C}^{\otimes n}_{\vec{t}}$ up to a given error. }
\end{figure}
 
The channel communication task studied in this  paper  is  depicted in Figure \ref{fig-visible}.   A client chooses  a quantum channel $\map{C}_\vec{t}$ from a parametric family   $\{\map{C}_\vec{t}\}_{\vec{t}\in\set{T}} \subseteq {\sf Chan}  (\spc H_{\rm in},\spc H_{\rm out})$, where  $\vec t$ is a vector of parameters and  $\set T$ is a suitable manifold, which we take to be  a bounded subset of $\R^v$ for some $v\in\N$. The client encodes the description of the   channel $\map{C}_\vec{t}$ into  a quantum state $\eta_{\vec{t},n} \in {\sf St} (\spc H_{{\rm prog},n})$ of a given finite-dimensional quantum system, called the {\em program system}.     Then, the client sends    
the program system to the server via a noiseless quantum communication link, and the server uses a decoding channel $\map D_n \in  {\sf Chan}  (   \spc H_{{\rm prog},n}\otimes\spc H_{\rm in}^{\otimes n}   \, ,\,  \spc  H_{\rm out}^{\otimes n}     )$ to execute an approximation of the channel  $\map{C}_\vec{t}^{\otimes n}$.  

The state $\eta_{\vec{t},  n}$ will be called  a  {\em quantum program} for the channel $\map{C}_\vec{t}^{\otimes n}$.   A channel communication protocol is specified by a triple ${\sf P}_n:  = (  d_{{\rm prog}, n},  \{ \eta_{\vec{t}, n }\}  ,  \map D_n)$, where $d_{{\rm prog}, n}=  \dim (\spc H_{{\rm prog}, n})$ is the dimension of the program system, $\{ \eta_{\vec{t}, n }\} \subseteq {\sf St} (\spc H_{{\rm prog},n})$ are the program states, and $\map D_n$ is the decoding channel.

 The performance of a channel communication protocol can be measured by  the distance between the desired channel $\map{C}_\vec{t}^{\otimes n}$ and the channel $\widetilde {\map C}_{\vec{t}, n}$ implemented by the server on the target systems. Explicitly, the action of the channel $\widetilde {\map C}_{\vec{t}, n}$  is given by
\begin{align}
\widetilde {\map C}_{\vec{t}, n}  (\rho)  :  = \map D_n  ( \eta_{\vec{t},  n} \otimes \rho)   \qquad \forall \rho \in  L(\spc H_{\rm in}^{\otimes n}) \, .
\end{align}
Its deviation from the desired channel $\map{C}_\vec{t}^{\otimes n}$ will be measured by the diamond norm 
\begin{align}
&\left \|   \widetilde {\map C}_{\vec{t}, n}    - \map{C}_\vec{t}^{\otimes n} \right\|_\diamond  
\nonumber \\
:  =&   \sup_{  \psi   \in  {\sf PurSt} \left(\spc H_{{\rm in}}^{\otimes n}  \otimes \spc  H_{\rm in}^{\otimes n}\right)}      \quad   \left\| \,   \left[  \left(  \widetilde {\map C}_{{\vec{t}}, n}    - \map{C}_{\vec{t}}^{\otimes n} \right)  \otimes \map{I}_{\rm in}^{\otimes n}  \right]   \,\Big(\psi \Big)\right\|_1   \,  , 
\label{error-diamond}
\end{align} 
where  $\map{I}_{\rm in} \in {\sf Map}  (\spc H_{\rm in})$ is the identity map, and $\|  A \|_1  :=  \Tr{  \sqrt {A^\dag A}}$ is the trace-norm of a generic linear operator $A$. 
The error of a channel communication protocol ${\sf  P}_n$ is defined as 
\begin{align}\label{progerrn}
  \epsilon_{{\rm prog}}  ({\sf P}_n)    :  =  \sup_{\vec t \in\set T}     \left \|   \widetilde {\map C}_{\vec{t}, n}    - \map{C}_\vec{t}^{\otimes n} \right\|_\diamond      \, .
\end{align}
 For a sequence of channel communication protocols 
$\{  {\sf P}_n\}_{n\in \N}$,  we define the  error as 
\begin{align}\label{limitcommerror}
\epsilon_{\rm prog}  (\{  {\sf P}_n\}) :  = \limsup_{n\to \infty}    \epsilon_{{\rm prog}}   (  {\sf P_n} )  \, .
\end{align} 
If the  error satisfies the condition $\epsilon_{\rm prog}  (\{  {\sf P}_n\})   \le  \epsilon_{\rm prog}$ for some non-negative number $\epsilon_{\rm prog}$, then we say that  the sequence has error threshold $\epsilon_{\rm prog}$.   If $\epsilon_{\rm prog}$ is zero, we say that the sequence of protocols  $\{{\sf P_n}\}$ has asymptotically vanishing error. 

Given an error threshold  $\epsilon_{\rm prog}$, the goal is to minimize the dimension of the program system needed to achieve such error threshold.  The dimension of the program system   determines the {\em communication cost}, that is, the number of qubits that have to be transmitted from the client to the server. Explicitly, the communication cost is given by  $\log d_{{\rm prog},n}$ where $\log$ denotes the base-2 logarithm.  

For a sequence of channel communication protocols $\{  {\sf P}_n\}_{n\in \N}$ we will define the {\em regularized communication cost} 
\begin{align}
\gamma (  \{{\sf P_n}\})    : =  \limsup_{n\to \infty}  \frac {  \log  d_{{\rm prog},  n}}{\log n} \, .     
\end{align}
The reason for dividing by $\log n$ will become clear later in the paper, where we will show that the leading order of the communication cost scales as $\log n$. 

The infimum of $\gamma (  \{{\sf P_n}\})$ over all  possible sequences of channel communication protocols with error threshold $\epsilon_{\rm prog}$ will be called the {\em regularized communication cost with error threshold $\epsilon_{\rm prog}$}.    



For $n=1$, our channel communication task coincides with the task of programming quantum channels  \cite{nielsen1997programmable,  fiuravsek2002universal, bergou2005universal, hillery2006approximate, d2005efficient,ishizaka2008asymptotic,sedlak2019optimal,kubicki2019resource}, and has  been recently used for  studying  communication capacities \cite{pirandola2017fundamental,das2017quantum}.    Our channel communication task can be described as the task of programming the channels $\{\map C_{\vec t}^{\otimes n}\}$. We will focus on the large $n$ limit, studying how the communication cost grows with $n$.  


\section{Remote channel simulation}\label{sec-simulation}

A lower bound on the communication cost of a family of channels can be obtained from a more general task, which we call {\em remote channel simulation}.  

\begin{figure}  [h!]
\begin{center}
  \includegraphics[width=0.8\linewidth]{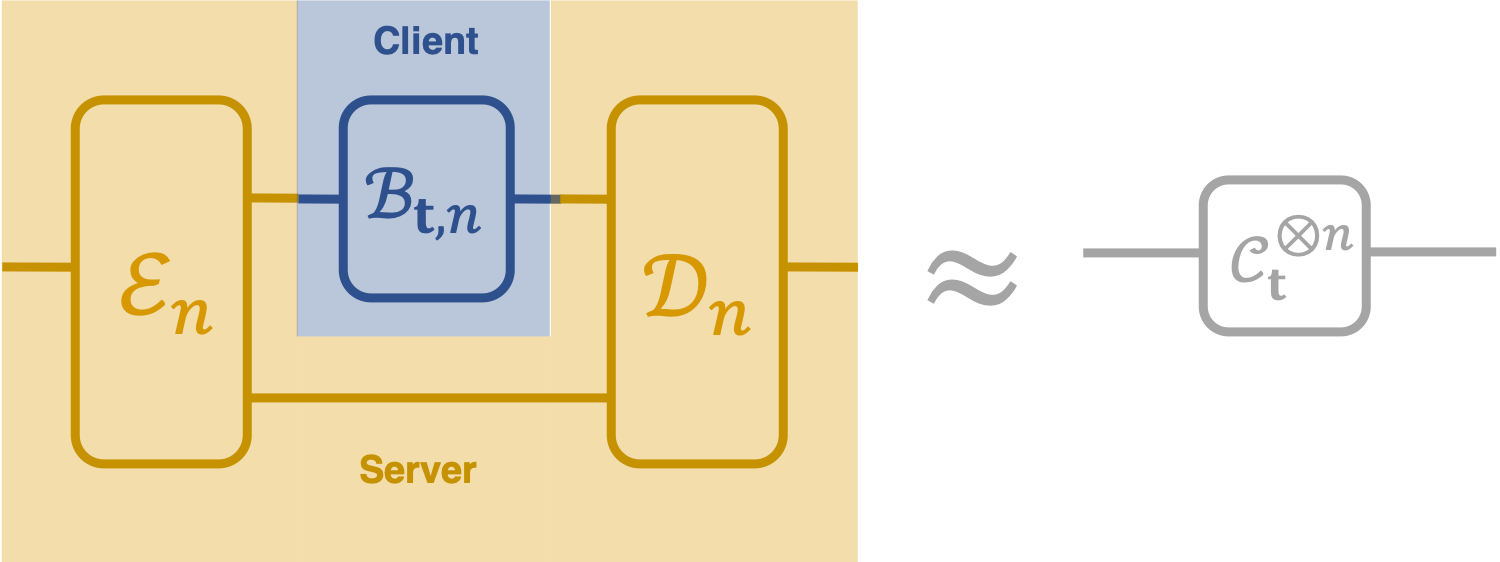}
  \end{center}
\caption{\label{fig-simulation}
  {\bf Remote channel simulation.}   A client, equipped with a small quantum computer, executes a channel  $\map{B}_{\vec{t},n}$, which enables a server to  reproduce $n$ uses of a target channel  $\map C_{\vec t}$.  The part in gold, i.e.\ the channels $\map{E}_n$ and $\map{D}_n$, are the encoder and the decoder of used by the server to interact with the client, while the part in blue, i.e. the channel $\map{B}_{\vec{t},n}$, is the channel performed by the client.   The protocol is designed so that the overall transformation at the server's end approximates the target channel $\map C_{\vec{t}}^{\otimes n}$  for every choice of $\vec{t}$ in a given parameter set $\set{T}$. }
\end{figure}

The remote channel simulation task is   depicted in Figure \ref{fig-simulation}.   A client, equipped with a small quantum computer, wants a server to execute $n$ parallel uses of a quantum channel $\map C_{\vec{t}}$, chosen from a parametric family $\{\map{C}_\vec{t}\}_{\vec{t}\in\set{T}}$.   To this purpose, the server encodes the input  of the desired channel $\map C_{\vec{t}}^{\otimes n}$ using  a quantum channel    $\map E_n  \in  {\sf Chan}  (\spc H_{\rm in}^{\otimes n},    \spc H_{{\rm in},n} \otimes \spc M_{n})$, called the {\em encoder}. The encoder outputs two systems:  a small system with Hilbert space $\spc H_{{\rm in},n}$, which is sent to the client through a noiseless quantum communication link, and a larger system with Hilbert space $\spc M_{n}$, which is stored in a quantum memory at the server's end.  Then, the client uses its quantum computer to  execute a channel  $\map{B}_{\vec{t}, n} \in  {\sf Chan}  (\spc H_{{\rm in},n},  \spc H_{{\rm out},n} )$, producing a  small output   system with Hilbert space $\spc H_{{\rm out}, n}$. After the channel  $\map{B}_{\vec{t},n}$  has acted, the client  sends the output system to the server. Finally, the server applies  a decoder $\map D_n \in  {\sf Chan}  (\spc H_{{\rm out}, n} \otimes \spc M_n,  \spc H_{\rm out}^{\otimes n}) $. The protocol is designed in such a way that the overall transformation  $  \widetilde{\map C}_{\vec{t},  n} :  = \map D_n  (\map B_{\vec{t}, n}  \otimes \map I_{\spc M_n }) \map E_n$ is close to the desired channel  $\map C_{\vec t}^{\otimes n}$.  

A remote channel simulation protocol is described by a quadruple ${\sf P_n} :  =  \{  d_{{\rm in},n}  , d_{{\rm out},n} ,\map E_n, \map D_n \}$.   As an error measure,  we will use the diamond norm $  \|   \widetilde{\map C}_{\vec{t}  ,  n}   -  \map C_{\vec{t}}^{\otimes n}\|_{\diamond}$.   For a remote channel simulation protocol ${\sf P}_n$, we define the  simulation error as \begin{align}\label{simerrn}
\epsilon_{{\rm sim}}  ({\sf P}_n)  :  = \sup_{\vec t\in\set T}  \|   \widetilde{\map C}_{\vec{t}  ,  n}   -  \map C_{\vec{t}}^{\otimes n}\|_{\diamond} \, .
\end{align}  For a sequence of remote channel simulation protocols 
$\{  {\sf P}_n\}_{n\in \N}$, we define the simulation error as $\epsilon_{\rm sim} (  \{  {\sf P}_n \})  :  = \limsup_{n\to \infty}    \epsilon_{{\rm sim}}  (  {\sf P}_n)$. If the error satisfies the condition $\epsilon_{\rm sim} (  \{  {\sf P}_n \})  \le \epsilon_{\rm sim}$ for some non-negative number $\epsilon_{\rm sim}$, then  we say that  the sequence $\{  {\sf P}_n\}$ has error threshold $\epsilon_{\rm sim}$.    If $\epsilon_{\rm sim}$ is zero, we say that the sequence of protocols  $\{{\sf P_n}\}$ has asymptotically vanishing error.

To quantify the communication cost of remote channel simulation, we will use the total number of qubits exchanged between the client and the server, namely $Q_{\rm tot} (n) : = \log d_{{\rm in},n}  +  \log d_{{\rm out},n}$, where $d_{{\rm in},  n}$ and $d_{{\rm out},n}$ are  the dimensions of the input and output systems of channel $\map B_{\vec{t},  n}$.     
For a sequence of remote channel simulation protocols 
$\{  {\sf P}_n\}_{n\in \N}$
 we will define the {\em regularized simulation cost} 
\begin{align}
\sigma (  \{{\sf P_n}\})    : =  \limsup_{n\to \infty}  \frac {    Q_{\rm tot}  (n)}{\log n} \, .     
\end{align}
The reason for dividing by $\log n$ will become clear later in the paper, where we will show that the total  number of qubits  $Q_{\rm tot} (n)$ grows as $\log n$ at the leading order.  

The infimum of $\sigma (  \{{\sf P_n}\})$ over all  possible sequences of remote channel simulation  protocols with error threshold $\epsilon_{\rm sim}$ will be called the {\em regularized simulation cost with error threshold $\epsilon_{\rm sim}$}.

The insertion of a quantum channel between an encoder and a decoder,  using a quantum memory as in Figure \ref{fig-simulation}, represents the most general transformation from quantum channels to quantum channels. Such transformations are known as a {\em quantum supermap} \cite{chiribella2008transforming,chiribella2009theoretical,chiribella2013quantum}.  In our setting, the supermap implemented by the client and server represents the ``transmission of a quantum channel'' from the client to the server.   We stress  that the quantum channel  $\map C_{\vec t}^{\otimes n}$ implemented in the protocol  is {\em not a  communication channel between the client and the server}: here, the term ``quantum channel"  is used as a  technical term for  a generic quantum process.

 We also stress  that  channel simulation task considered in this paper is different from the task considered in  the quantum reverse Shannon theorem \cite{bennett2002entanglement,berta2010conceptually,berta2011quantum,datta2012quantum,bennett2014quantum}, which sometimes is also referred to as ``channel simulation''.  The quantum reverse Shannon theorem concerns   the entanglement-assisted classical communication cost of converting an identity channel  into a desired  channel, in a scenario where the input of the channel is held by one party, and the output of the channel is held by the other party. In contrast, our channel simulation problem  concerns the total  (classical and quantum) communication cost   in a scenario where the input and output of the simulating channel are held by one party, while the input and output of the simulated channel are held by the other party.    

  The channel communication task described in Section \ref{sec:chancommun} is a special case of the channel simulation where the input of the channel $\map B_{\vec{t}, n}$ is trivial,  and therefore the channel $\map B_{\vec{t},  n}$ is simply the preparation of a state $\eta_{\vec{t}, n}$. Hence, every lower bound on the total amount of communication required in the channel simulation scenario is also a lower bound on the amount of communication required  in the channel communication scenario.

\section{Basic notions from quantum metrology } \label{sec-metrology}

 Quantum metrology 
\cite{holevo,helstrom,giovannetti2006quantum,giovannetti2011advances,giovannetti2006quantum,luis1996optimum,buvzek1999optimal,chiribella2004efficient,bagan2004quantum,hayashi2006parallel} investigates the use of quantum resources, such as entanglement and coherence, to enhance the precision of parameter estimation.

\begin{figure}  [b!]
\begin{center}
  \includegraphics[width=0.6\linewidth]{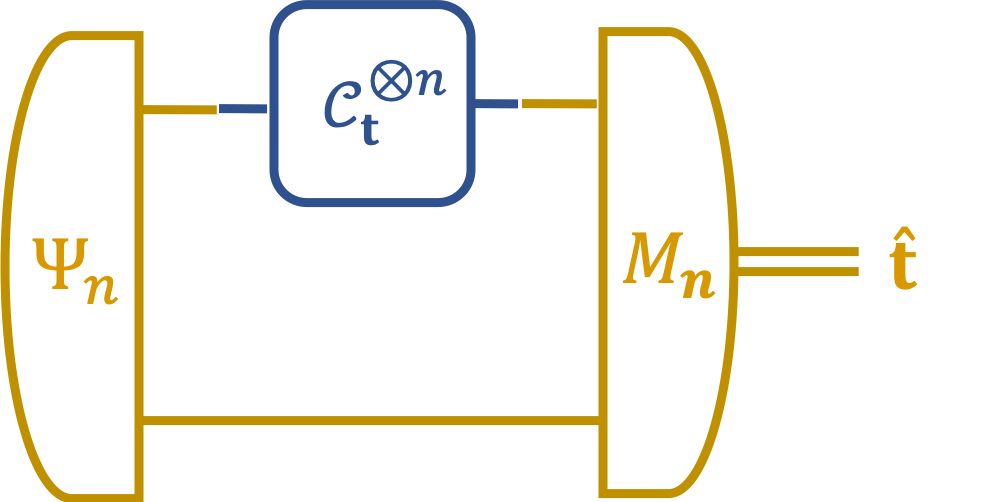}
  \end{center}
\caption{\label{fig-metrology}
  {\bf Quantum metrology scheme.}  A typical scenario in quantum metrology concerns estimation of the channel $\map C_{\vec{t}}$, parametrised by unknown parameters $\vec{t}$, from $n$ parallel uses. An experimenter prepares a probe state $\psi_n$ and sends part of it through $\map{C}_{\vec{t}}^{\otimes n}$, keeping the remaining part in a memory. The experimenter then measures jointly the output of $\map{C}_{\vec{t}}^{\otimes n}$ and the state of the memory with a quantum measurement, which, combined with classical data-processing, yields an estimate $\hat{\vec{t}}$ of $\vec{t}$.  The measurement and the classical data-processing are jointly described by a POVM $M_n  (\d  \hat {\vec t})$. }
\end{figure}

In the typical setting, an experimenter is given access to $n$ parallel uses of a channel $\map{C}_{\vec{t}}$ that depends on some unknown parameters $\vec{t} \in \set T\subset  \R^v$. The experimenter then probes the channel in order to obtain an  estimate $\hat{\vec{t}}$ of the parameters $\vec{t}$.  The estimation setup consists of the following recipe, illustrated in Figure \ref{fig-metrology}: 
\begin{enumerate}
\item Prepare a state $\psi_n\in\operatorname{St}\left(\spc{H}_{\rm in}^{\otimes n}\otimes\spc{R}_{~n}\right)$, where $\spc{H}_{\rm in}$ is the input Hilbert space of $\map{C}_{\vec{t}}$ and ${\spc R}_{~n}$ is the Hilbert space of a {\em reference system}, unaffected by the action of the channel.
\item Send the $n$ input systems   through $\map{C}_{\vec{t}}^{\otimes n}$.
\item Measure the $n$ output systems together with the reference.
\item Use the measurement outcome to produce an estimate $\hat{\vec t}$. 
\end{enumerate}
 The last two steps can be concisely represented by using a positive operator-valued measure (POVM) $M_n   (\d\hat{ \vec t})$, which incorporates both  the experimental  setup used to measure the system and  the classical algorithm that produces the estimate $\hat{\vec t}$ from the measurement outcome.

The error of the  estimation procedure can be quantified in terms of the mean squared error (MSE) matrix, defined as 
\begin{align}\label{covmat}
&V_{ij}\big(\psi_n,\map{C}^{\otimes n}_\vec{t}, M_{n }  ( \d \hat{\vec{t}} ) \big)
\nonumber \\
:=& \int   \,    (t_i-\hat{t}_i)(t_j-\hat{t}_j)  \, \Tr\left[M_n  (\d \hat{\vec{t}}) \, \map{C}^{\otimes n}_\vec{t}(\psi_n)\right] \, .
\end{align}  
In the following, we will refer to the trace of the MSE  matrix as to the {\em mean squared error (MSE)}, denoted as 
\begin{align}
{\sf MSE}  \big(\psi_n,\map{C}^{\otimes n}_\vec{t}, M_n (\d \hat{\vec{t}}  )  \big):= \Tr  \Big[V  \big(\psi_n,\map{C}^{\otimes n}_\vec{t}, M_n (\d \hat{\vec{t}} ) \big)  \Big]
\end{align}

Note that the MSE depends on the value of the true parameter $\vec t$, which is unknown to the experimenter.   To  account for this fact, we  consider the worst-case over all possible values of $\vec t$.  
Minimizing the  worst-case MSE  over all choices of $\psi_n$ and $M_n  (   \d \hat{\vec{t}})  $,  we obtain the quantity  
\begin{align}\label{MSE}
 {\sf MSE}      (\{\map{C}^{\otimes n}_\vec{t}\}   ):  =\min_{\psi, \, M_{n}  (\d \hat{\vec{t}})}  \, \sup_{\vec t \in\set T}  ~ {\sf MSE}  \big(\psi_n,\map{C}^{\otimes n}_\vec{t},  M_n(\d \hat{\vec{t}})  \big)\, ,
 \end{align} 
which we call the MSE of the channel family $\{\map C_{\vec t}^{\otimes n}\}_{\vec t\in\set T}$.

The MSE of the channel family $\{\map{C}_{\vec{t}}^{\otimes n}\}$ vanishes as a function of $n$, at a rate that  depends on  the family. 
For many channel families, the  MSE scales as $1/n$, a   scaling  known as the {\em standard quantum limit} \cite{giovannetti2006quantum}.  
As an example of families with this scaling,
we can list families of constant channels, that is, channels of the form $\map C_{\vec t}   (\rho)  =  \rho_{\vec t}$ for some state $\rho_{\vec t}$ independent of $\rho$.
For unitary channels, of the form $\map C_{\vec t}  (\rho)  =  U_{\vec t}  \rho  U_{\vec t}^\dag$ for some unitary operator $U_{\vec t}$, the MSE scales as $1/n^2$ \cite{buvzek1999optimal,chiribella2004efficient,bagan2004quantum,hayashi2006parallel,kahn2007fast}.  The $(1/n^2)$-scaling is referred to as the {\em Heisenberg  limit} \cite{giovannetti2006quantum}.

We say that the channel family  $\{\map{C}^{\otimes n}_\vec{t}\}_{\vec{t}\in\set{T}}$ is {\em Heisenberg limited}  (respectively, {\em standard quantum limited}) if its MSE scales as
\begin{align}
{\sf MSE}\left(\{\map{C}^{\otimes n}_\vec{t}\}\right)=\Theta\left(n^{-2}\right)\,\left({\rm respectively,}~ \Theta\left(n^{-1}\right)\right)  \, ,
\end{align}
where  the notation $g(n)=\Theta\left(f(n)\right)$ means that $g(n)$ and $f(n)$ have the same asymptotic behavior for large enough $n$. More precisely, there exists an integer $n_0$ and two constants $c_1$ and $c_2$ such that $c_1\cdot f(n)\le g(n)\le c_2\cdot f(n)$ for every $n>n_0$.

An important tool for deriving bounds on the precision of quantum metrology is the quantum Fisher information matrix, or more precisely, the quantum Fisher information matrices, for there exist multiple quantum versions  of the classical Fisher information matrix.    In this paper we will mostly use the quantum Fisher information matrix defined in terms of the {\em right logarithmic derivatives (RLDs)}.  For a $C^1$-continuous family of quantum states $\{\rho_{\vec t}\}_{t\in \set T}$,  the RLDs are the operators $\{L^{\rm R}_i\}_{i=1}^{v}$ defined via the equation   $\partial\rho_{\vec t}/\partial t_i=\rho_{\vec t}L^{\rm R}_i$.  The RLD quantum Fisher information matrix  is then defined as  
\begin{align} 
J^{\rm R} (\rho_{\vec t})  :  =   \left[\Tr\left(\left(L_i^{\rm R}\right)^\dag\frac{\partial \rho_{\vec{t}}}{\partial t_j}\right)\right]_{ij}
 \end{align}   

For a $C^1$-continuous family of quantum channels $\{  \map C_{\vec t}\} \subseteq {\sf Chan}  (\spc H_{\rm in}, \spc H_{\rm out})$, we  define the {\em maximum  RLD Fisher information norm}  as
	\begin{align}
	\label{JR} 
	 J^{\rm R}_{\map{C}_{\vec{t}}}:=\max_{\psi\in\operatorname{St}(\spc{H}_{\rm in}\otimes\spc H_{\rm in})} \left\|J^{\rm R}\Big((\map{C}_{\vec{t}}\otimes\map{I}_{\rm in}  )  \,  (\psi)\Big) \right\|_{\infty} \, ,
		\end{align} 
		where $\|  M \|_{\infty}  :  = \sup_{ \vec s  \in  \R^v,  \|  \vec s \|=1}  \, \|  M  \vec s \|$ denotes the Schatten $\infty$-norm of a generic  real $v\times v$ matrix $M$, $M\vec s \in \R^v$ is the vector with components $(M \vec s)_i   =  \sum_j   M_{ij}   \, s_j$,  and $\|  s\|   :  = \sqrt{    \sum_{i=1}^v  s_i^2 }$ is the Euclidean norm. 
			
When the channels $\{  \map C_{\vec t}\}$ satisfy a suitable condition, the RLD Fisher information norm can be computed through a simple analytical expression. The condition is expressed in terms of the Choi operator $C_{\vec t}  :  =  \Choi (\map C_{\vec t})$ and reads as follows:  
\begin{cond}\label{cond:derivative1}  For every vector $\vec s  \in  \R^v$, the support of the directional derivative  $\left( \partial  C_{\vec t  +  x\, \vec s} / \partial x \right)_{x=0}$ is contained in the support of $C_{\vec t}$. 
\end{cond}

When Condition \ref{cond:derivative1} is satisfied, the RLD Fisher information norm has the following expression:

\begin{prop}\label{lem:explicit}
If   the channel family $\{ \map C_{\vec t}\}_{\vec t \in \set T}$ satisfies Condition \ref{cond:derivative1} at point $\vec t$,  then its RLD Fisher information norm at $\vec t$ is given by 
\begin{align}\label{RLD1}
J^{\rm R}_{\map{C}_{\vec{t}}}   =    \max_{  \vec s  \in \R^v, \,   \|  \vec s\|=1}     \left\|   \Tr_{\rm out}  
\left[     \Big(   \frac{  \partial    C_{\vec t  +  x  \,  \vec s}}{\partial x}\Big)_{x=0}    C_{\vec t}^{-1}   \Big(  \frac{  \partial    C_{\vec t  +  x  \, \vec s}}{\partial x}\Big)_{x=0}         \right]\right\|_\infty \, .
\end{align} 
\end{prop}

The proof is provided in Appendix \ref{app:RLD}. 

We note that every channel family satisfying Condition \ref{cond:derivative1} is necessarily standard quantum limited, as proven in Appendix \ref{app:SQL}.   The converse is not true: there exist channel families that are standard quantum limited, and yet violate Condition \ref{cond:derivative1}. For example, consider a family consisting of constant channels of the form $\map C_{\vec t} (\cdot) :  =  \Tr[\cdot]  \,  \psi_{\vec t}$, where $\{\psi_{\vec t}\} \subseteq {\sf PurSt}  (\spc H_{\rm out})$ is a family of pure states.  
In this case, the support of the Choi operator $C_{\vec t}$ is $\Span  \{|\psi_{\vec t}  \>\} \otimes \spc H_{\rm in}$, while the support of its directional derivative is $\Span  \{\partial  |\psi_{{\vec t}  + x \, {\vec s} }  \>/\partial x\} \otimes \spc H_{\rm in}$, and the two supports are different for some direction $\vec s$, except  in the trivial case where the pure state $\psi_{\vec t}$ is independent of $\vec t$.

\section{Main results} 

Here we summarize the main results of the paper.  The standing assumptions for these results are  that the channel family under consideration is $C^1$-continuous, and that the parameter manifold $\set T$ is the Cartesian product of $v$ intervals, namely $\set{T}=  \set{T}_1 \times \cdots \times \set {T}_v$,  with $\set{T}_i=[t_{0,i},t_{1,i}]$ for $i  \in \{1,\dots, v\}$.

The first result is a lower bound on the communication cost of remote channel simulation protocols (cf. Figure \ref{fig-simulation}).  The cost of remote channel simulation was determined by Fang et al.\ \cite{fang2019quantum} in terms of the channel's maximum output mutual information.   Our result   establishes a connection between the simulation cost and the precision limits of quantum metrology.

\begin{theo}[Quantum metrology  lower bound on remote channel simulation]\label{theo-converse}
If the channels $\{\map{C}^{\otimes n}_\vec{t}\}_{\vec{t}\in\set{T}}$  can be estimated with MSE  $O(n^{-\beta})$, then their regularized simulation cost with error threshold $\epsilon_{{\rm sim}}$ is lower bounded by $(1-\epsilon_{{\rm sim}}) \, (v\beta/2) $, where $v$ is the dimension of the parameter manifold $\set{T}$.   
In particular, the regularized simulation cost is lower bounded by $v\beta/2$ for all sequences of channel simulation protocols with asymptotically vanishing error. 
\end{theo} 

Theorem \ref{theo-converse} implies   that the simulation cost of standard quantum limited  (respectively, Heisenberg limited) channels grows    at least $   (1-\epsilon') \, v/2 \, \log n$  (respectively, $(1-\epsilon')  v\log n$) for every   $\epsilon' >  \epsilon_{\rm sim}$.   More generally, any intermediate MSE scaling $n^{-\beta}$ will result in a simulation cost of  at least $(1-\epsilon')  \, (v\beta/2 )\,  \log n$ at the leading order in $n$.  

Theorem \ref{theo-converse} also yields a lower bound on the communication cost of quantum channels. Indeed, a protocol that communicates $\map{C}_{\vec{t}}^{\otimes n}$ from the server to the client can  be regarded as a special case of remote simulation protocol, with $d_{{\rm in},n}=1$. This argument leads to  the following corollary:
\begin{cor}\label{cor-program}
If the quantum channels  $\{\map{C}^{\otimes n}_\vec{t}\}_{\vec{t}\in\set{T}}$  can be estimated with MSE $O(n^{-\beta})$, then their regularized communication cost with error threshold $\epsilon_{\rm prog}$  is lower bounded by $(1-\epsilon_{{\rm prog}}) \, (v\beta/2) $.    In particular, the regularized communication n cost is lower bounded by $v\beta/2$ for all sequences of channel communication protocols with asymptotically vanishing error. 
\end{cor}

Corollary \ref{cor-program} can be applied to  the problem of programming quantum channels. 
  A set of quantum channels $\{\map{C}_\vec{t}\}_{\vec{t}\in\set{T}}$ is called {\it programmable} if there exist a set of quantum states $\{\eta_\vec{t}\}_{\vec{t}\in\set{T}}$ and a quantum channel $\map{W}$ so that
$\map{C}_\vec{t}(\rho)=\map{W}(\rho\otimes \eta_\vec{t})$ holds for every $\vec{t}\in\set{T}$.
 It is known that  programmable quantum channels are standard quantum limited
 \cite{ji2008parameter,pirandola2017ultimate,erratumarticle,takeoka2016optimal}. 
Corollary \ref{cor-program} implies that the communication cost of programmable channels is at least $  (1 -\epsilon')  v/2\,   \log n$ at the leading order in $n$, for every $\epsilon' >  \epsilon_{\rm sim}$.


 Our second main result is  an achievability result.  We provide an explicit  channel communication protocol working for a special class of standard quantum limited channels. The protocol is ``classical'', in the sense that it can be implemented using only a classical communication channel between the client and the server.   The main idea is that the client and the server choose a suitable discretization of the parameter manifold, and the client communicates to the server the point in the discretization that is closest to the label of the desired channel.

  The protocol  works for channel families  $\{\map{C}_\vec{t}\}_{\vec{t}\in\set{T}}$  
satisfying the following conditions: 
\begin{cond}\label{cond:RLD}
The supremum  of the  RLD Fisher information norm over the channel family is non-zero, {\em i.~e.}\ $\sup_{\vec{t} \in \set{T}}
J^{\rm R}_{\map{C}_{\vec{t}}}>0$.
   	\end{cond}
\begin{cond}\label{cond:samesupport} The support of  the Choi operator $\Choi ({\map{C}_{\vec{t}}})$ is independent of $\vec{t}$.   
\end{cond}  

 Condition \ref{cond:RLD} is rather weak: it simply states that the channel family $\{\map C_{\vec t}\}$ has a non-trivial dependence on $\vec t$.  {In a sense, Condition \ref{cond:samesupport} is also rather weak, in that it is satisfied by all the channels in the interior of the convex set of quantum channels. Such channels have Choi operators with full rank, and therefore satisfy Condition \ref{cond:samesupport}. Hence, any quantum channel is arbitrarily close to a channel satisfying Condition \ref{cond:samesupport}. On the other hand, Condition \ref{cond:samesupport} rules out some interesting families of quantum channels, such as the families of unitary channels.   It is easy to see that } Condition \ref{cond:samesupport} is stronger than Condition \ref{cond:derivative1} in Section \ref{sec-metrology}.  As a consequence,  a  channel family  $\{\map{C}^{\otimes n}_{\vec t}\}$ satisfying Condition \ref{cond:samesupport}  is  standard quantum limited, cf. Appendix \ref{app:SQL}.  The channels satisfying Condition \ref{cond:samesupport} form a strict subset of the standard quantum limited channels: for example, constant channels of the form $\map C_{\vec t}  (\rho)  =  |\psi_{\vec t}\>\<\psi_{\vec t}|$ are standard quantum limited, but the support of their Choi operator  depends on $\vec t$.    
Still, the set of channel families satisfying Condition \ref{cond:samesupport}   is large enough to contain interesting examples.  For example,  it contains all  families of  Pauli channels  of the form $\map{C}_{\vec{t}}:=  p_{0 ,\vec t}  \, \map I  +  p_{x,\vec t}  \, \map X  +  p_{y,\vec t}  \, \map Y  +   p_{z,\vec t}  \, \map Z $, where $\map X, \map Y,\map Z$ are the unitary channels corresponding to the three Pauli matrices, and 
  $p_{\vec t}  =  ( p_{0,\vec t},  p_{x,\vec t}, p_{y,\vec t}, p_{z,\vec t} )$ is a probability distribution with support independent of $\vec t$.

The details of the channel  communication protocols are as follows: 
\begin{algorithm}[H]
  \caption{Communicating  $n$ uses of channel in a family $\{\map{C}^{\otimes n}_\vec{t}\}_{\vec{t}\in\set{T}}$ satisfying Conditions \ref{cond:RLD} and \ref{cond:samesupport}.}
  \label{protocol-classical}
   \begin{algorithmic}[1]
 
\Statex (Preparation) 
The client and the server use  a  discretization  of  
$\set{T}$, denoted by $\set{T}_{n}$ and defined as
\begin{align}\label{discret-set}
\set{T}_{n}:=
\Bigg\{ 
\frac{n^{-\alpha-\frac12}}{\sqrt{v J^{\rm R}_{\max}}}
\vec{z} \Bigg| \vec{z} \in \mathbb{Z}^v 
\Bigg\} \cap \set{T} \, ,\quad{\rm with} \quad J^{\rm R}_{\max} := \sup_{\vec{t} \in \set{T}}
J^{\rm R}_{\map{C}_{\vec{t}}} \,.
\end{align}
  Note  that the discretization $\set{T}_{n}$ is well-defined, thanks to  Condition \ref{cond:RLD}.  
\State (Encoding.) To communicate the  channel $\map{C}_{\vec{t}}^{\otimes n}$,  
the client finds an element  $\vec{t}_{n}\in\set{T}_{n}$ that satisfies
   	\begin{align}\label{tr}
	&\|\vec{t}_{n}-\vec{t}\|< 
	\frac{n^{-\alpha-\frac12}}{\sqrt{J^{\rm R}_{\max}}}
	\end{align} 
	in terms of the Euclidean distance.   Note that such $\vec{t}_{n}$ always exists thanks to Eq.\ (\ref{discret-set}).  The client encodes the discretized vector $\vec{t}_{n}$ into a pure  quantum state  $\eta_{\vec{t},n}    =  |\vec{t}_n\>\<\vec{t}_n|$, where $\{ |\vec{t}_n\>\}$ is an orthonormal basis for the Hilbert space of the program system.  
	\State (Transmission)     The client sends the program system to the server. 
\State (Decoding.) The server implements $\map{C}_{\vec{t}_n}^{\otimes n}$ on the input state via the following measure-and-operate decoder
\begin{align}
\map{D}(A\otimes B):=\sum_{\vec{t}_{n}\in\set{T}_{n}}\Tr[|\vec{t}_{n}\>\<\vec{t}_{n}|A]\,\map{C}_{\vec{t}_n}^{\otimes n}(B),
\end{align}
where $A$ and $B$ are arbitrary operators on the Hilbert space of the program and the Hilbert space of the input, respectively.
 \end{algorithmic}
\end{algorithm}


The cost and the error rate of the above protocol are quantified in the following theorem, whose proof  is provided  in Section \ref{sec-proof-direct}. 
\begin{theo}\label{theo-HL}
Under Conditions \ref{cond:RLD} and \ref{cond:samesupport}, Protocol \ref{protocol-classical} has an error $O(n^{-\alpha})$ and costs 
$(1/2+\alpha)v\log n$ bits of communication at the leading order in $n$, where  $v$ is the dimension of the parameter manifold $\set{T}$ and  $\alpha  >0$ is an arbitrary positive number.  
\end{theo}

Theorem \ref{theo-HL}, combined with Corollary \ref{cor-program} implies the optimality of  Protocol \ref{protocol-classical}  for  communicating channels satisfying Conditions \ref{cond:RLD} and \ref{cond:samesupport} in the case of asymptotically vanishing error.  Indeed,  the infimum of the regularized communication cost over all channel communication protocols with  $\epsilon_{\rm prog}=0$ is exactly $v/2 \log n$, matching the lower bound in Corollary \ref{cor-program}.

\section{Proof of Theorem \ref{theo-converse}}\label{sec-converse}
\subsection{Preliminaries on the inaccuracy of state estimation}
The proof of Theorem \ref{theo-converse} is based on the notion of inaccuracy of  an \red{estimate} of a state family \cite{yang2018quantum}.   
In the following, we will first review a few basic facts about the inaccuracy in state estimation
 and then we will provide  the proof of Theorem \ref{theo-converse}. 

For the estimation of a family of quantum states $\{    \rho_{\vec t}\}_{\vec t\in\set T}$,  the estimate $\hat{\vec t}$ is generated by a POVM $\{  M ( \d \hat{\vec t})  \}$, and its \red{conditional} 
probability distribution is $p(\d \hat{\vec t}|  \vec t)   = \Tr [  M (\d  \hat {\vec t}) \,  \rho_{\vec t}] $.    We regard the  estimate $\hat {\vec t}$ as the value of a random variable $\widehat{\vec T}$, distributed with probability $p(\d \hat{\vec t}|  \vec t)   $.  The random variable $\widehat{\vec T}$ will also be called the {\em estimate} of the parameter $\vec t$.   

The inaccuracy of the estimate $\widehat{\vec T}$    quantifies the radius of the smallest Euclidean ball in which the \red{estimate} has confidence $p\in(0,1)$.  
Explicitly, the inaccuracy of the \red{estimate} $\widehat {\vec T}$ at the true value $\vec t$ is defined as \cite{yang2018quantum}
\begin{align}
\delta (p,  \vec t, \widehat {\vec T})   :  =  \inf  \Big\{\delta \in \R  ~\Big|~   \red{\mathbf{Pr}_{\widehat{\vec T}|\vec t} } 
\left[  \|  \widehat{\vec T}-\vec t\|\le \delta\right]\ge  p\Big\} ,\label{H616}
\end{align}
where \red{$\mathbf{Pr}_{\widehat{\vec T}|\vec t}  
\left[  \|  \widehat{\vec T}-\vec t\|\le \delta\right]$  
is the conditional} probability that the random variable $\widehat{\vec T}$ takes a value   
  $\hat{\vec t}$ such that the Euclidean distance $\|  \hat{\vec t}  - \vec t\| $ is no larger than $\delta$. 

For an arbitrary \red{estimate}, the  inaccuracy is related to the MSE  by the following inequality 
\begin{align}\label{Chebyshev2}
\delta (p,\vec t,\widehat {\vec T})      \le    \sqrt{ \frac{ {\sf  MSE}   (\vec t,  \widehat {\vec T})}{ 1-p}}     \qquad \forall p\in(0,1) \,,   \forall \vec t\in\set T \, ,
\end{align}
where  ${\sf MSE}({\vec t},  \widehat {\vec T})  :  =  \sum_{i=1}^v   \int p( \d  \hat  {\vec t}| \vec t)  \,     (  \hat t_i  - t_i )^2$ is the MSE of the \red{estimate} 
$ \widehat {\vec T}$  at the true value $\vec t$ (see Appendix \ref{app:Chebyshev} for a proof).  We remind the readers that $\hat{\vec T}$ in the above definition is a random variable, which takes value $\hat{\vec t}$ with probability $p( \d  \hat  {\vec t}| \vec t)$.

\red{For another  state family $\{    \rho_{\vec t}'\}_{\vec t\in\set T}$, 
we have another conditional distribution 
$p'(\d \hat {\vec t}|\vec t):  = \Tr [  M(\d \hat {\vec t}) \, \rho'_{\vec  t}]$.
In the following, we denote the variable subject to the above conditional distribution 
by $\widehat  {\vec T}'$, i.e., $\mathbf{Pr}_{\widehat{\vec T}'|\vec t}(\d \hat {\vec t})=p'(\d \hat {\vec t}|\vec t)$.
Hence, we denote the inaccuracy defined \eqref{H616} with the conditional distribution $\mathbf{Pr}_{\widehat{\vec T}'|\vec t}$
by $\delta (p,\vec t,\widehat {\vec T}')$.}
If two quantum states  $\rho_{\vec t}$ and $\rho_{\vec t}'$ satisfy the condition $\|\rho_{\vec t}'  -\rho_{\vec t}   \|_1<  \epsilon$, then the above inaccuracies 
$\delta (p,\vec t,\widehat {\vec T})$ and $\delta  (p,\vec t,\widehat {\vec T}')$ obey  the continuity property  \cite[Eq. (7.1)]{yang2018quantum}
\begin{align}\label{continuity}
\delta (p -\epsilon,  \vec t  ,\widehat {\vec T}) \le 
\delta (p,  \vec t,\widehat {\vec T}')  \le \delta (  p+\epsilon,\vec t,\widehat {\vec T}) \, .
\end{align}

The inaccuracy is also related to the mutual information between the estimate $\hat{\vec t}$ and the true parameter $\vec t$.  
To establish the relation, one has to assign a prior  probability distribution to the true parameter $\vec t$. Denoting by $\hat {\vec T}$ the corresponding random variable, one has the following
\begin{lem}\label{lemma-distortion}
Let   $\vec T$ be a   $v$-dimensional continuous   random variable with domain $\set T  = \prod_{i=1}^v  [  t_{0,i},  t_{1,i}]$,   and let  $\widehat {\vec T}$ be an   \red{estimate} \red{subject to a conditional distribution $\mathbf{Pr}_{\widehat{\vec T}|\vec t} $ when $\vec T$ takes value $\vec t$.}
We define the worst-case inaccuracy 
$\delta_p:  =  \sup_{\vec t \in \set T}   \delta (p, \vec t,\widehat {\vec T}) $, the volume $|\set T|   :=  \prod_{i=1}^v   (  t_{1,i } -  t_{0,i} )$ of the domain, 
and the volume 
$B_{v,\delta_p}:=\frac{(\sqrt{\pi}\delta_p)^v}{\Gamma(v/2+1)}$  
of a $v$-dimensional Euclidean ball with radius $\delta_p$, 
where $\Gamma(y)$ denotes the Gamma function.   
When the condition 
\begin{align}
\log\left(\frac{1-p }{p}\right)
\le \log\left(\frac{2^v|\set{T}|-B_{v,\delta_p}}{B_{v,\delta_p}}\right) 
 \label{H614}
\end{align}
holds,
the mutual information between $\widehat {\vec T}$ and $\vec T$, denoted by $I(\widehat{\vec T}:{\vec T})$,  satisfies the bound
\begin{align}
& I(\widehat{\vec T}:{\vec T})\nonumber \\
\ge & H({\vec T})-p\log\left(\frac{B_{v,\delta_p}}{p}\right)-(1-p)\log\left(\frac{2^v  |\set T| -B_{v,\delta_p}}{1-p}\right) \, ,
\label{mutualinfo1}
\end{align}
where   $H({\vec T})$ denotes the differential entropy of the random variable $\vec T$.
The bound can be further weakened to 
\begin{align}
I(\widehat{\vec T}:{\vec T})  \ge
& H(\vec T) -pv\log(\sqrt{\pi}\delta_p)+p\log\Gamma\left(\frac v2+1\right)   \nonumber \\
&-(1-p)\log(2^v|\set{T}|)-h(p) \, .
\label{mutualinfo2}
\end{align} 
\end{lem}
The proof is provided in Appendix \ref{A1}.


\begin{figure}  [t!]
\begin{center}
  \includegraphics[width=0.6\linewidth]{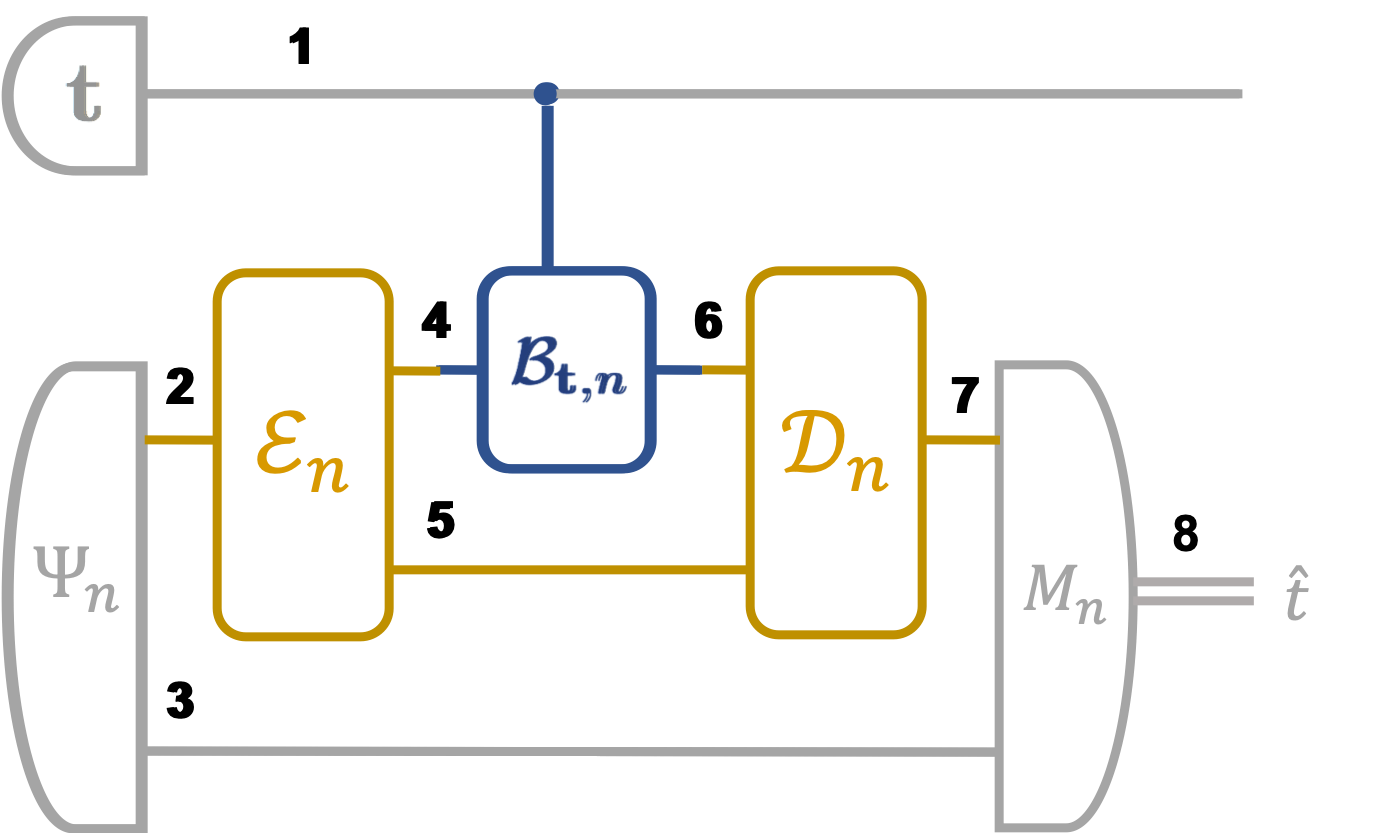}
  \end{center}
\caption{\label{fig-distortion}
  {\bf Parameter estimation in the remote channel simulation scenario.}  
  The estimation of the channel $\map C_{\vec{t}}$ from $n$ parallel uses is (approximately) related to the estimation of the channel $\map B_{\vec{t} ,n}$ acting on lower-dimensional quantum systems.   In particular, every protocol for estimating $\map C_{\vec{t}}$ from $n$ parallel uses can be used as a protocol for estimating channel $\map B_{\vec{t} ,n}$, by replacing the channel $\map C_{\vec t}^{\otimes n}$ with the channel $\widetilde {\map C}_{\vec t, n}:  =   \map D_n (  \map B_{\vec t, n}\otimes \map I_{\spc M})  \map E_n$ resulting from a remote channel simulation protocol.     The figure shows the general form of such a protocol, consisting in transforming the channel $\map B_{\vec{t} ,n}$ into (an approximation of) channel $\map C_{\vec{t}}^{\otimes n}$, by using an encoder $\map E_n$, and a decoder $\map D_n$.  The parameter $\vec t$ is then estimated by preparing an input state $\psi_n$ for (the approximation of) channel $\map C_{\vec{t}}^{\otimes n}$,  and implementing a POVM  $M_n  (\d \hat{\vec{t}}  )$ on the output. }
\end{figure}  



\subsection{Proof of Theorem \ref{theo-converse}} 
For the  whole proof, $p$ will denote  an arbitrary number in the open interval 
$(0,1-\epsilon_{{\rm sim}})$.
By  assumption, the parametric family  $\{\map{C}^{\otimes n}_\vec{t}\}_{\vec{t}\in\set{T}}$  can be estimated with MSE  $O(n^{-\beta})$.
This means that, for sufficiently large $n$,  there exist a reference system $\spc R_{~n}$, 
an input state $\psi_n\in {\sf St} (  \spc H_{\rm in}^{\otimes n}  \otimes \spc R_{~n})$,  and a POVM  $M_n  (\d \hat{\vec{t}}  ) $ such that the MSE    satisfies the bound
\begin{align}\label{fgh}
{\sf MSE} \big(\psi_n,\map{C}^{\otimes n}_\vec{t},  M_n (\d \hat{\vec{t}}  ) \big)\le c\cdot n^{-\beta}  \qquad \forall \vec t \in \set T\, ,
\end{align}
where  $c>0$ is a suitable constant,  and 
 ${\sf MSE}\big(\psi_n,\map{C}^{\otimes n}_\vec{t}, M_n  (\d \hat{\vec{t}}  )  \big)$
is the mean squared error defined  in Eq. (\ref{MSE}). 

In this protocol, the output states $   \rho_{{\vec t}, n }   :  =   ( \map C_{\vec t}^{\otimes n}\otimes \map I_{{\map R}_n}   ) (\psi_n)$ are measured with the POVM $M_n  (\d \hat{\vec t})$.  Let $\widehat {\vec T}_n$ be the \red{estimate} with 
\red{the conditional} probability distribution $p_n (\d\hat {\vec t}|  \vec t):  =  \Tr [  M_n  (\d \hat {\vec t}) \,  \rho_{{\vec t}, n}]$,   and let  
 $\delta  (p,\vec t,\widehat {\vec T}_n)$ be the corresponding inaccuracy defined in Eq.  \eqref{H616}.
By inserting Eq. (\ref{fgh})   into the bound  (\ref{Chebyshev2}), we obtain the bound
\begin{align}
\delta (p,  \vec t,\widehat {\vec T}_n)\le   \sqrt{\frac{c}{(1-p)n^{\beta}}} 
\qquad   \forall \vec t\in\set T
\label{H2}.
\end{align}

Now,  let $\{   {\sf P}_n\}_{n\in \N}$ be an arbitrary sequence of remote simulation protocols  with error threshold    $\epsilon_{\rm sim}$.     For each protocol ${\sf P}_n=  (d_{{\rm in},  n} ,   d_{{\rm out},  n},  \map E_n, \map D_n)$, consider  the channel simulation $\widetilde{\map C}_{\vec{t},  n} :  = \map D_n  (\map B_{\vec{t}, n}  \otimes \map I_{\spc M_n}) \map E_n$, and let $\epsilon_{{\rm sim},  n}  :  = \epsilon_{\rm sim}   ({\sf P}_n)$ be the error defined in Eq. (\ref{simerrn}).   Since the error threshold satisfies the condition  $\epsilon_{\rm sim}   \ge    \epsilon (\{{\sf P}_n\}) : =   \limsup_{n\to \infty}  \epsilon_{{\rm sim},  n}$, and since $p$ satisfies the condition $p <  1-\epsilon_{\rm sim}$,  there exists an integer $n_0$ such that the condition  $  p <  1-\epsilon_{{\rm sim},n} $ holds for every $n\ge n_0$.  
 
By construction,  the output states $   \rho_{{\vec t}, n} $
and $\rho_{{\vec t},n}'  :  = ( \widetilde{\map C}_{{\vec t},n}\otimes \map I_{\map R}   ) (\psi)$ are at most $\epsilon_{{\rm sim},  n}$ apart from each other  in trace norm.
Let us denote by $ \widehat {\vec T}_n'$  
\red{the variable with the conditional probability distribution 
$\mathbf{Pr}_{\widehat{\vec T}_n'|\vec t}(d \hat{\vec t}) $ given by}
$p_n' (  \d \hat {\vec t} | \vec t) :  =  \Tr [ M_n (\d \hat {\vec t})\,     \rho_{{\vec t }, n}']$, and by     $\delta  (p,\vec t,\widehat {\vec T}_n')$ the corresponding inaccuracy defined in \eqref{H616}.
The continuity of the inaccuracy (\ref{continuity}) yields the bound 
\begin{align}
\delta(p, \vec t,\widehat {\vec T}_n')  
\le   \delta (  p  +  \epsilon_{{\rm sim},  n},  \vec t,\widehat {\vec T}_n) 
\le  \sqrt{\frac{c}{(1-p-\epsilon_{{\rm sim},n})n^{\beta}}}  
\end{align}
valid for every 
$ \vec t\in\set T$
and
$n\ge n_0$. 
Since the bound holds for every $\vec t\in\set T$,  it also holds for the worst-case inaccuracy 
$\delta_{p,n}:  =  \max_{\vec t \in \set T} \delta (p, \vec t,\widehat {\vec T}_n') $,
which satisfies the inequality 
\begin{align}\label{zxc}
\delta_{p,n}' \le 
\sqrt{\frac{c}{(1-p-\epsilon_{{\rm sim},n})n^{\beta}}} \, .
\end{align}
 
We now use the relation between the inaccuracy and the mutual information  (Lemma \ref{lemma-distortion}).  Referring to Figure \ref{fig-distortion} for the labelling of the Hilbert spaces, we have that the total number of transmitted qubits satisfies the   bound 
\begin{align}
\nonumber Q_{\rm tot} (n)&:=\log d_{4}+\log d_6  \ge H(4)+H(6)\\
&=H(3,5)+H(6)
\ge H(3,5,6)\nonumber\\
&\ge I(1:3,5,6)
\ge I(1:8)   \equiv  I(  \widehat {\vec  T}_n'   :   \vec T)\, ,\label{inter1}
\end{align}
where $d_k$ is the dimension of the Hilbert space of system $k $,  $H(k)$ is the von Neumann entropy of the reduced state of  system  $k\in \{1,\dots ,8\}$, $H(3,5)$  (respectively, $H(3,5,6)$) is  the von Neumann entropy of the composite system made of systems $3$ and $5$  (respectively, $3$, $5$, and $6$),  $I(1:3,5,6)$ is the mutual information between system $1$ and the composite system made of systems $3$, $5$, and $6$, and $I(  1;8 )$ is the mutual information between systems $1$ and $8$. 
The first inequality comes from the maximum value of the von Neumann entropy.  The first equality holds since the encoder of the client, i.e. $\map{E}_A$, can be assumed w.l.o.g.\ to be isometric. Indeed, for any protocol with a non-isometric $\map{E}_A$, one can construct a protocol in which $\map{E}_A$ is replaced by its Stinespring dilation with its purifying system traced out at the decoder stage. Such a protocol would have the same error and the same communication cost as the original protocol. The second inequality is the subadditivity of the von Neumann entropy. The third inequality holds since the system labeled by $1$ is a classical system, and therefore the conditional entropy is non-negative. The fourth inequality is the data processing inequality of the mutual information.


\red{Now, for $p>0$ and for large enough $n$, the condition \eqref{H614} holds
with the conditional distribution $\mathbf{Pr}_{\widehat{\vec T}_n'|\vec t} $:
indeed, 
for large enough $n$,
$\delta_{p,n}'$ is close to zero (due to Eq. \eqref{zxc}), which implies  that
$B_{v,\delta_{p,n}'}$ is sufficiently small compared to $2^v|\set{T}|$, and therefore the condition
$\frac{2^v|\set{T}|-B_{v,\delta_{p,n}'}}{B_{v,\delta_{p,n}'}}\ge \frac{1-p}{p}$ holds, thus implying $\log \frac{2^v|\set{T}|-B_{v,\delta_{p,n}'}}{B_{v,\delta_p'}} \ge 
\log \frac{1-p}{p}$. }

Since condition  \eqref{H614} holds,  we can apply Lemma \ref{lemma-distortion}
\red{to the conditional distribution $\mathbf{Pr}_{\widehat{\vec T}_n'|\vec t} $}.
Using  Eq. (\ref{mutualinfo2}), we then get
\begin{align}
I(  \widehat {\vec  T}_n'   :   \vec T)
\ge &  H(\vec T) -pv\log\left(\sqrt{\pi}\,
\delta_{p,n}'   \right)-(1-p)\log\big(2^v \, |\set T| \big)\nonumber \\
&+p\log\Gamma\left(\frac v2+1\right) -h(p)\, .
\end{align}
Since the distribution of $\vec{t}$ is arbitrary, we can choose it to be uniform, and therefore with entropy  $H(\vec T)=\log|\set{T}|$. The above inequality becomes
\begin{align}\label{inter2}
I(  \widehat {\vec  T}_n'   :   \vec T)\ge 
&-pv\log\left(\sqrt{\pi}\,
\delta_{p,n}'\right)-(1-p)v \nonumber \\
&+p\log\left(|\set{T}|\Gamma\left(\frac v2+1\right)\right)-h(p).
\end{align}
Combining Eq. (\ref{inter2}) and Eq. (\ref{inter1}), one gets that 
\begin{align}
Q_{\rm tot} (n)\ge & 
-pv\log\left(\sqrt{\pi}\,
\delta_{p,n}'
\right)-  (1-p)v \nonumber \\
& +p\log\left(|\set{T}|\Gamma\left(\frac v2+1\right)\right)-h(p).
\label{inter3}
\end{align}
Inserting Eq. (\ref{zxc}) into the above bound, we finally obtain
\begin{align}
Q_{\rm tot}  (n)  \ge &- pv \log \sqrt{\pi \,c/[(1-p-\epsilon_{{\rm sim},n})n^{\beta}]}  -(1-p)v
\nonumber \\
&+p\log\left(|\set{T}|\Gamma\left(\frac v2+1\right)\right)-h(p)\nonumber \\
\ge &- pv \log  \sqrt{\pi \,c/[(1-p-\epsilon_{{\rm sim},n})n^{\beta}]} 
-(1-p)v\nonumber \\
&+p\log\left(|\set{T}|\Gamma\left(\frac v2+1\right)\right)-h(p)\nonumber \\
=& 
 \frac{p v  \beta}{2}\, \log n
 - pv \log  \sqrt{\pi \,c/[(1-p-\epsilon_{{\rm sim},n})]} \nonumber \\
&-(1-p)v+p\log\left(|\set{T}|\Gamma\left(\frac v2+1\right)\right)-h(p) \, .
\end{align}

Dividing both sides by $\log n$ and taking the limit superior, we obtain 
\begin{align}
\limsup_{n\to \infty}\frac{Q_{\rm tot}  (n)}{\log n }  \ge \frac{p v  \beta}{2}.
\end{align}
Since $p$ is an arbitrary real number in $(0,1-\epsilon_{{\rm sim}})$,
the  statement in Theorem \ref{theo-converse} follows by taking the supremum over $p$. 
\qed

\medskip

The quantity on the left hand side of Eq. (\ref{inter2}) is the mutual information between the true value $\vec{t}$ and its estimate $\hat{\vec{t}}$ in the setting of channel estimation, which  amounts to the number of digits of $\vec{t}$ that can be specified  in quantum metrology \cite{hassani2017digital}.   When the channel is Heisenberg limited, it is immediate from Eq. (\ref{inter2}) that this quantity scales as $\log n$, which was called the ``information theoretic Heisenberg limit'' in  \cite{hassani2017digital}, which focussed on the case of ideal phase estimation. Our Eq. (\ref{inter2}) establishes a general lower bound on the digitization of estimation precision, which extends the result in \cite{hassani2017digital} from ideal phase estimation  to the general noisy and multi-parameter metrology.

\section{Proof of Theorem \ref{theo-HL}}\label{sec-proof-direct}
\subsection{Preliminaries on the RLD Fisher information norm and the $2$-R\'{e}nyi divergence}
The proof uses a relation between the RLD Fisher information norm and the  $2$-R\'{e}nyi divergence    \cite{sharma2013fundamental,wilde2014strong,Leditzky2018,fang2019geometric}.

The $2$-R\'{e}nyi divergence  for two states $\rho$ and $\sigma$ is defined as $D_2(\rho||\sigma)=\log\Tr [\rho^2\sigma^{-1}]$ for $\operatorname{supp}(\rho)\subset\operatorname{supp}(\sigma)$ \cite{sharma2013fundamental,wilde2014strong}.  
The $2$-R\'{e}nyi divergence for states can be extended to a $2$-R\'{e}nyi divergence for quantum channels in the straightforward  way: for two channels $\map A$ and $\map B$, the $2$-R\'{e}nyi divergence is defined by applying them to an input state (possibly including a reference system), and by maximizing the $2$-R\'{e}nyi divergence of the resulting output states. Explicitly, one has   
\begin{align}
&D_2(\map{A}||\map{B})
\nonumber \\
:= &
\sup_{\psi\in\operatorname{PurSt}(\spc{H}_{\rm in}\otimes \spc{H}_{\rm in})} \quad D_2\Big( \,  (\map{A}\otimes\map{I}_{\rm in}) \,(\psi)  \Big \| \,  (\map{B}\otimes\map{I}_{\rm in} )\, (\psi)\Big).
\label{def-D2}
\end{align}
The  $2$-R\'{e}nyi divergence of quantum channels is  an example of generalized divergence for channels, in the sense of  \cite[Definition II.2]{Leditzky2018}. Properties of the $\alpha$-R\'{e}nyi divergence for $\alpha \ge 0$  have been studied in \cite{fang2019geometric}. 

Locally, the  $2$-R\'{e}nyi divergence is related to the RLD Fisher information norm by the following relation:   

\begin{lem}\label{lem:infinitesimal}
Let $\{  \map C_{\vec t}\}_{\vec t \in \set T}$ be a parametric family of quantum channels with the property that all the corresponding  Choi operators have the same support. Then, the 2-R\'{e}nyi divergence has the Taylor expansion 
\begin{align}\label{expansion}
D_2\left(\map{C}_{\vec{t}'}||\map{C}_{\vec{t}}\right)&
\le \|\vec{t}'-\vec{t}\|^2  J^{\rm R}_{\map{C}_{\vec{t}}}
+O\left(\|\vec{t}'-\vec{t}\|^3\right) 
\end{align}
for every $  \vec t \in  \set T  $ and $\vec t' \in \set T$,
where $J^{\rm R}_{\map{C}_{\vec{t}}}$ is the RLD  Fisher information norm
defined in Eq. (\ref{JR}).   
\end{lem}

The proof of Lemma \ref{lem:infinitesimal} is provided in Appendix \ref{app:2Renyi}.

\subsection{Proof of Theorem \ref{theo-HL}}
It is not hard to see that the  Protocol  \ref{protocol-classical}  demands $(1/2+\alpha)v\log n$ bits of communication at the leading order of $n$. Indeed,  the communication cost is the cost of transmitting an element in the set $\set{T}_{n}$ [cf.\ Eq.\ (\ref{discret-set})], which can be upper bounded as
\begin{align}
\log |\set{T}_{n}|\le \log\prod_{i=1}^{v}\left(\frac{ t_{1,i}  -  t_{0,i}}{n^{-\alpha-\frac12}/\sqrt{v J^{\rm R}_{\max }}}+1\right) \, .
\end{align}
 It is clear from the above bound that the communication cost is $(1/2+\alpha)v\log n$ bits at the leading order of $n$.



We now show that the error vanishes as $n^{-\alpha}$.  Recall that the error is  $\epsilon_{{\rm prog}, n}  :  =\sup_{\vec t \in \set T}  \epsilon_{{\rm prog}, n} (\vec t)$ with $\epsilon_{{\rm prog}, n}  (\vec t) : =    \left \|  \map{C}_{\vec{t}_{n}}^{\otimes n}     -  \map{C}_{\vec{t}}^{\otimes n} \right\|_{\diamond}$.    Let  $\psi^{\rm wc}_{\vec{t}} \in  \operatorname{ PurSt}  \left(\spc H_{\rm in}^{\otimes n} \otimes \spc H_{\rm in}^{\otimes n}  \right)$ be the input state such that 
\begin{align}
 &\left \|  \map{C}_{\vec{t}_{n}}^{\otimes n}     -  \map{C}_{\vec{t}}^{\otimes n} \right\|_{\diamond}  \nonumber \\
 =&  \left\|  \,(\map{C}_{\vec{t}_{n}}^{\otimes n}\otimes\map{I}_{\rm in}^{\otimes n}) \, (\psi^{\rm wc}_{\vec{t}})   -   \,(\map{C}_{\vec{t}}^{\otimes n}\otimes\map{I}_{\rm in}^{\otimes n}) \, (\psi^{\rm wc}_{\vec{t}})  \right\|_1 \, .
\end{align}
Applying Pinsker's inequality \cite{ohya2004quantum}, one has
\begin{align*}
&\epsilon_{{\rm prog}, n}  (\vec t) \\
\le &\sqrt{\frac{2}{\log e }D_1\Big(  (\map{C}_{\vec{t}_{n}}^{\otimes n}\otimes\map{I}_{\rm in}^{\otimes n})(\psi^{\rm wc}_{\vec{t}}) \Big\|  (\map{C}_{\vec{t}}^{\otimes n}\otimes \map{I}_{\rm in}^{\otimes n})(\psi^{\rm wc}_{\vec{t}})\Big)} \, ,
\end{align*}
 where $D_1  ( \rho  \| \sigma) :  =  \lim_{\alpha  \to 1}  D_\alpha  ( \rho  \| \sigma)  =\Tr [ \rho  (\log \rho  -  \log \sigma)]$ is the quantum relative entropy, and coincides  with the $\alpha$-R\'{e}nyi divergence in the limit $\alpha \to 1$.    Using the monotonicity of the $\alpha$-R\'{e}nyi divergence with respect to $\alpha$ \cite{muller2013quantum}, we then obtain the bound 
\begin{align}
&\epsilon_{{\rm prog},  n}  (\vec t) \nonumber \\
\le & \sqrt{\frac{2}{\log e}D_2\Big((\map{C}_{\vec{t}_{n}}^{\otimes n}\otimes\map{I}_{\rm in}^{\otimes n})(\psi^{\rm wc}_{\vec{t}})  \Big \| (\map{C}_{\vec{t}}^{\otimes n}\otimes \map{I}_{\rm in}^{\otimes n})(\psi^{\rm wc}_{\vec{t}})\Big)} \, .
\label{poi}
\end{align}

By definition, of the $2$-R\'{e}nyi divergence for channels, one has 
\begin{align}
&D_2\Big(     \,(\map{C}_{\vec{t}_{n}}^{\otimes n}\otimes\map{I}_{\rm in}^{\otimes n}) \, (\psi^{\rm wc}_{\vec{t}}) \Big\|  \,  (\map{C}_{\vec{t}}^{\otimes n}\otimes \map{I}_{\rm in}^{\otimes n}) \, (\psi^{\rm wc}_{\vec{t}})\Big) \nonumber \\
\le &D_2\Big(\map{C}^{\otimes n}_{\vec{t}_{n}} \Big\| \map{C}^{\otimes n}_{\vec{t}}\Big).
\label{int2}
\end{align}  
Now, we use the fact that the $2$-R\'{e}nyi divergence  is additive    (see \cite[Item 3 of Theorem 3]{fang2019geometric} and Corollary \ref{cor:additivity} in Appendix \ref{app:2Renyi}), meaning that we have 
\begin{align}\label{int1}
D_2\left(\map{C}^{\otimes n}_{\vec{t}_{n}}||\map{C}^{\otimes n}_{\vec{t}}\right)=n \, D_2\left(\map{C}_{\vec{t}_{n}}||\map{C}_{\vec{t}}\right).
\end{align}

Inserting Eqs.  (\ref{int2}) and (\ref{int1})  into Eq. (\ref{poi}), we obtain
\begin{align}\label{int3}
\epsilon_{{\rm prog}, n} (\vec t)\le\sqrt{\frac{2n}{\log e}D_2\left(\map{C}_{\vec{t}_{n}}||\map{C}_{\vec{t}}\right)}.
\end{align}

Finally, we express the 2-Renyi divergence in terms of the RLD Fisher information norm, using Lemma \ref{lem:infinitesimal}.  Inserting  Eq.\ (\ref{expansion}) into Eq. (\ref{int3}),  we obtain the bound 
\begin{align}
\nonumber \epsilon_{{\rm prog}, n}   (\vec t)    &\le \sqrt{\frac{2n}{\log e}  \,  \left[\|\vec{t}_{n}-\vec{t}\|^2  J^{\rm R}_{\map{C}_{\vec{t}}}
+O\left(\|\vec{t}_{n}-\vec{t}\|^3\right)  \right]}  \\
 \nonumber  &   
  \le \sqrt{\frac{2n}{\log e}  \,  \left[\frac{n^{-2\alpha-1} }{ J^{\rm R}_{\max}}    \,   J^{\rm R}_{\map{C}_{\vec{t}}}  +O\left(\|\vec{t}_{n}-\vec{t}\|^3\right)\right] } \\
 \nonumber  &  \le  \sqrt{ \frac 2 {\log e}  \left[  n^{-2\alpha}
+O(n ^{-3 \alpha-1/2})\right]
}\\
  &  =   \sqrt{\frac{ 2 }{ \log e}} \,   n^{-\alpha}   +    O\left(  n^{-  2\alpha-1/2}\right) \, ,
\end{align} 
where the second inequality follows from Eq.\ (\ref{tr}). 
Since the bound holds for every $\vec t$, we have proven the error vanishes as $n^{-\alpha}$.  \qed

\section{Conclusion} 
We studied the cost of communicating $n$ parallel uses of an unknown quantum channel, chosen from a given parametric family. 
In the direct part, we proposed a protocol for sending the classical description of the channel.
In the converse part, we derived a lower bound for the more general task of remote channel simulation, where a client, equipped with a small quantum computer, enables a server to  execute a desired quantum channel on a large quantum system.   The bound on remote channel simulation  yields the desired bound for communicating quantum channels as a corollary.
The bound is achieved by our concrete protocol for channels satisfying certain conditions.
The bound captures the  measurement sensitivity of quantum channels from an information-theoretic point of view and is therefore a step towards the unification of quantum metrology and quantum Shannon theory \cite{hassani2017digital,czajkowski2017super,yang2018quantum}.
Potentially, the bound may have applications in various directions of delegate quantum computation  \cite{childs2005secure}, where a server is asked to execute a computation on the state held by a remote client.
It can also, for example, be used to determine the bandwidth of a quantum sensor network \cite{komar2014quantum} and to hint on how quantum programs can be conceived.
These applications will become more desired as quantum devices are assembled into a network in the near future.

An interesting problem for future research  is the compression of multiple-use channels, where the goal is to encode $n$ uses of an unknown quantum channel $\map C_{\vec{t}}$  into another quantum channel  $\map B_{\vec{t},n}$ acting on a smaller system.  This task is very similar to the simulation task considered in this paper, except that  the parameter $\vec{t}$ is  now {\em invisible}.  
 The counterpart of this task for states is the task of compressing multicopy states, recently studied both theoretically \cite{plesch2010efficient,yang2016efficient,yang2016optimal,yang2018compression,yang2018quantum} and experimentally \cite{rozema2014quantum}.   The task of channel compression is more involved since the input  of the channel is not necessary in the many-copy form, and these compression protocols for states cannot be applied directly.  Our bound on remote channel simulation (cf.\ Theorem \ref{theo-converse}) applies also to compression, since it is harder. However, it remains open whether a concrete protocol achieving the bound exists.


\appendices

\section{Proof of Proposition \ref{lem:explicit}}\label{app:RLD} 
In the scalar case $v=1$,  Eq. (\ref{RLD1}) was shown in   \cite[Theorem 1]{hayashi2011comparison}. 
The extension to the multi-parameter case $v>1$ can be derived as follows. First,  note that one has
\begin{align}
\nonumber 
&   J^{\rm R}_{\map C_{\vec t}}     \\
   =&  \max_{\psi  \in  \operatorname {PurSt}  (\spc H_{\rm in} \otimes \spc H_{\rm in})}  \,      \max_{\vec s  \in  \R^v \, ,\| \vec s  \|=  1}    \sum_{i,j} \,      s_i     \,\left[  J^{\rm R}  \Big (  \,( \map C_{\vec t} \otimes \map I)   ( \psi) \right]_{ij}  \,  s_j    \nonumber\\
\nonumber =&        \max_{\vec s  \in  \R^v \, ,\| \vec s  \|=  1}  \,  \max_{\psi  \in  \operatorname {PurSt}  (\spc H_{\rm in} \otimes \spc H_{\rm in})}  \,    \sum_{i,j} \,      s_i     \,\left[  J^{\rm R}  \Big (  \,( \map C_{\vec t} \otimes \map I)   ( \psi) \right]_{ij}  \,  s_j    \\
\nonumber =&    \max_{\vec s  \in  \R^v \, ,\| \vec s  \|=  1}      \,    \max_{\psi  \in  \operatorname {PurSt}  (\spc H_{\rm in} \otimes \spc H_{\rm in})}  \,    J^{\rm R}  \Big (  \,( \map D_{\vec s\, ,x} \otimes \map I)   ( \psi) \Big)\\
=  &  \max_{\vec s  \in  \R^v \, ,\| \vec s  \|=  1}      \,    J^{\rm R}_{ \map D_{\vec s\, ,x}}  \,,    \label{vvv} 
\end{align}
where $ J^{\rm R}  \Big (  \,( \map D_{\vec s\, ,x} \otimes \map I)   ( \psi) \Big) $ is the  RLD Fisher information for the one-parameter family $\{   ( \map D_{\vec s\, ,x} \otimes \map I)   ( \psi)  \}_{x}$,  with  $\map D_{\vec s, x}   :  =  \map C_{  \vec t   +x\,  \vec s  }$.   Then, \cite[Theorem 1]{hayashi2011comparison} guarantees the equality 
\begin{align}
 J^{\rm R}_{ \map D_{\vec s\, ,x}}  =    \left\|  \Tr_{\rm out}  \left[     \left(   \frac{  \partial    C_{\vec t  +  x  \,  \vec s}}{\partial x}\right)_{x=0}     \,  C_{\vec t}^{-1}  \, \left(  \frac{  \partial    C_{\vec t  +  x  \, \vec s}}{\partial x}\right)_{x=0}      \right]\right\|_\infty   \, ,
\end{align}
which, inserted in Eq. (\ref{vvv}) yields the desired result.    

\section{Proof that channels satisfying Condition \ref{cond:derivative1} are standard quantum limited}\label{app:SQL}  
To prove the desired result, we consider the quantity  
\begin{align}
&K (\{  \map C_{\vec t}\})  :  =\limsup_{n\to \infty} \, n  \,  {\sf MSE}  (\{  \map C_{\vec t}^{\otimes n}\})   
\nonumber \\
=&  \limsup_{n\to \infty}  \min_{\psi_n,   M_n  (\d \hat {\vec{t}})  }  \, \sup_{\vec t \in\set T}~    n  \,    {\sf MSE}  \left(\psi_n,\map{C}^{\otimes n}_{\vec t},M_n   (\d \hat{\vec{t}})  \right)  \, .    
\end{align}  
Note that  $K(\{\map{C}_\vec{t}\})$  is a non-zero constant if and only if the channels $\{\map C_{\vec t}\}$ are standard quantum limited. 

We now show that $K(\{\map{C}_\vec{t}\})$ is strictly positive for channel families satisfying Condition \ref{cond:derivative1}.

\begin{lem}
For a channel family $\{\map C_{\vec t}\}$ satisfying Condition \ref{cond:derivative1}, one has the bound
\begin{align}\label{SR5}
 K(\{\map{C}_{\vec{t}}\}) \ge  \frac{1}{J^{\rm R}_{\map{C}_{\vec{t}}} } .
		\end{align} 
\end{lem}

\noindent\Proof ~First of all, note that one has the bound 
\begin{align}
K(\{\map{C}_{\vec{t}}\})  \ge  K_{\rm la} (\map{C}_{\vec{t}_0})   \qquad\forall  t_0 \in\set T \,,
\end{align}  where $K_{\rm la} (\map{C}_{\vec{t}_0}) $ is  the {\em local asymptotic minimax risk}   \cite{hajek1972local}, defined as  
\begin{align}
&K_{\rm la}(\map{C}_{\vec{t}_0})
:=  \inf_{  \{(\psi_n,  M_n  (\d \hat {\vec t}))\}_{n\in\N} }  ~ \lim_{\epsilon\to 0}  ~  \nonumber \\
&\qquad\limsup_{n\to \infty} \Bigg[\sup_{ \vec t \in  U(  \vec t_0, \epsilon)}\, n  \, {\sf MSE}    \big(\psi_n,\map{C}^{\otimes n}_{\vec t}, M_n (\d \hat{\vec{t}})  \big) 
\Bigg],
\end{align}
where $\big\{\big(\psi_n,  M_n  (\d \hat {\vec t})\big)\big\}_{n\in\N} $  is a sequence of estimation strategies, consisting of a state $\psi_n$ and of a POVM $M_n  (\d \hat {\vec t})$, and $U(  \vec t_0, \epsilon)$ is the Euclidean ball of radius $\epsilon$ centred around $\vec t_0$. 
The key difference between  $K_{\rm la}(\map{C}_{\vec{t}_0})$ and $K(\map{C}_{\vec{t}})$ is that  $K_{\rm la}(\map{C}_{\vec{t}_0})$  focuses on the estimation error in a neighbourhood of a fixed $\vec{t}_0$, while $K(\map{C}_{\vec{t}})$ concerns the worst-case error over every $\vec{t}\in\set{T}$.

Let $\vec s \in R^v$ be an arbitrary unit vector.  By definition, one has   
\begin{align}
&K_{\rm la}(\map{C}_{\vec{t}_0})  \nonumber\\
  \ge &   \lim_{\epsilon\to 0}   \inf_{  \{(\psi_n,  M_n  (\d \hat {\vec t}))\}_{n\in\N} } \nonumber\\
  &\qquad  \limsup_{n\to \infty}  \Bigg[    \sup_{ \vec t \in  U(  \vec t_0, \epsilon)}\, n  \, {\sf MSE}    \big(\psi_n,\map{C}^{\otimes n}_{\vec t}, M_n (\d \hat{\vec{t}})  \big) \Bigg] \nonumber\\
  \ge &  \lim_{\epsilon\to 0}  \, \limsup_{n\to \infty}      \min_{\psi_n, \,  M_n (\d \hat{\vec{t}}  )  }   \, 
\sup_{\vec{t} \in U(\vec{t}_0,\epsilon)}      \,  n\,    \Tr \left[  V \big(\psi_n,\map{C}^{\otimes n}_{\vec t},   M_n (\d \hat{\vec{t}})  \big)\right]  \nonumber\\
  \ge & \lim_{\epsilon\to 0}  \, \limsup_{n\to \infty} 
        \nonumber \\
&\quad \min_{\psi_n, \,  M_n (\d \hat{\vec{t}} )  }   
 \Bigg[ \sup_{\vec{t} \in U(\vec{t}_0,\epsilon)}      \,  n\,   \left( \sum_{i,j}      s_i\,   \Big[   V \big(\psi_n,\map{C}^{\otimes n}_{\vec t},  M_n (\d \hat{\vec{t}})  \big) \Big]_{ij}  \, s_j \right)   \Bigg] \nonumber\\
 =& K_{\rm la}({\map D}_{\vec s,x})_{x=0} \,,  \label{abba}
\end{align} where   $\{\map D_{\vec s, x}\}_x$ is the one-parameter family defined by $\map D_{\vec s, x}  :  =  \map C_{\vec t_0 +  x\, \vec s}$.  

Since the family $\{\map C_{\vec t}\}$ satisfies Condition \ref{cond:derivative1} at $t$, the one-parameter family $\{\map D_{\vec s, x}\}$ satisfies Condition \ref{cond:derivative1} at $x=0$.  Hence,   \cite[Proposition 2]{hayashi2011comparison} guarantees the condition
       	\begin{align}
\label{SR3}
K_{\rm la}(\map{D}_{\vec s, x})_{x=0}=\lim_{n \to \infty}\frac{n}{J^{\rm S}_{{\map{D}}^{\otimes n}_{\vec s, x=0}}} \, ,
		\end{align} 
		where  ${J^{\rm S}_{{\map{D}}^{\otimes n}_{\vec s, x}}}$ is the quantum version of the Fisher information based on the symmetric logarithmic derivative (SLD).   We omit the definition of the SLD quantum Fisher information because it is not directly relevant here. What is relevant, instead, is the fact that the SLD quantum Fisher information is always upper bounded by the RLD quantum Fisher information  \cite{holevo}. Hence, one has the bound  
\begin{align}
J^{\rm S}_{{\map{D}}^{\otimes n}_{\vec s, x}}    \le    J^{\rm R}_{{\map{D}}^{\otimes n}_{\vec s, x}}   =  n \, J^{\rm R}_{{\map{D}}_{\vec s, x}},
\end{align} 
where the equality follows from the additivity of the RLD quantum  Fisher information  in the one-parameter case \cite[Corollary 1]{hayashi2011comparison}.   

Combining this inequality with Equations (\ref{abba}) and (\ref{SR3}), we obtain the bound 
    \begin{align}
 K(\{\map{C}_{\vec{t}}\}) \ge   \frac  1 {J^{\rm R}_{{\map{D}}_{\vec s, x=0}}}  \qquad \forall \vec s\in \R^v \,{\rm s.t.}\ \|  \vec s\|=1 \, . 
\end{align}
Since $J^{\rm R}_{{\map{C}}_{\vec t_0}}   =  \max_{\vec s}  \,  J^{\rm R}_{{\map{D}}_{\vec s, x=0}}$ (Equation (\ref{vvv})), this concludes the proof. \qed

\medskip 

In passing, we observe that the  inequality \eqref{SR5} yields  a necessary condition for the Heisenberg limit scaling (and more generally, for faster-than-standard-quantum-limit scalings): in order to have such scaling, the RLD Fisher information norm must be infinite.
This condition can be used to identify families of quantum channels beating the standard quantum limit. 

\section{Proof of Equation (\ref{Chebyshev2})}\label{app:Chebyshev}  

The proof follows the same steps of the proof of Chebyshev's inequality.  By definition, one has 
\begin{align}
&{\sf MSE}  (     \vec t ,  \widehat {\vec T})  
 = \int \d \hat {\vec t}  \,  \|   \hat {\vec t}   -  \vec t\|^2       \,  p(\hat{\vec t}| \vec t) \nonumber \\
 \nonumber     =& \int_{  \|   \hat {\vec t}  - \vec t \|  >\delta  } \d \hat {\vec t}  \,  \|   \hat {\vec t}   -  \vec t\|^2       \,  p(\hat{\vec t}| \vec t)   +   \int_{  \|   \hat {\vec t}  - \vec t \|  \le \delta  } \d \hat {\vec t}  \,  \|   \hat {\vec t}   -  \vec t\|^2       \,  p(\hat{\vec t}| \vec t)  \\
  \ge &  \delta^2  \,          \mathbf{Pr}  
\left(  \|  \widehat{\vec T}-\vec t\|> \delta\right) \, , 
\end{align}     
and therefore 
\begin{align}
\mathbf{Pr}  
\left(  \|  \widehat{\vec T}-\vec t\|> \delta\right)  \le  \frac {  {\sf MSE}  (     \vec t,  \widehat {\vec T} ) }{\delta^2} \, . 
\end{align}

Thanks to this inequality, every choice of $\delta$ satisfying the condition 
\begin{align}
\delta    >   \sqrt{\frac{{\sf MSE} (\vec t,  \widehat {\vec T})}{1-p}} 
\end{align}
will necessarily satisfy the condition $\mathbf{Pr}  \left[  \|  \widehat{\vec T}-\vec t\| >\delta \right] < 1-p$.     Hence, we obtain 
\begin{align}
&\nonumber \delta (p,  \vec t, \widehat{\vec T})  
  =  \inf  \Big\{\delta \in \R  ~\Big|~   \mathbf{Pr}  
\left[  \|  \widehat{\vec T}-\vec t\|\le \delta\right]\ge  p\Big\} \\
\le & \inf \left\{\delta \in \R ~\left|~  \delta    >   \sqrt{\frac{{\sf MSE} (\vec t,  \widehat {\vec T}) }{1-p}}    \right\}\right. 
= \sqrt{\frac{{\sf MSE} (\vec t,  \widehat {\vec T})}{1-p}}   \, .  \label{bastaa}
\end{align}

\section{Proof of Lemma \ref{lemma-distortion}}\label{A1}
\subsection{Preparation}
The proof of Lemma \ref{lemma-distortion} uses an auxiliary result, provided in the following:  
\begin{lem}
For every  \red{estimate} $\widehat{\vec T}$  of a given parameter $\vec t\in\set T$, the worst-case inaccuracy $\delta_p: = \sup_{\vec t \in\set T}   \delta (p,\vec t,\widehat{\vec T})$ satisfies the condition 
\begin{align}
\mathbf{Pr}\left[  \|\widehat{\vec T}-\vec t \|\le \delta_p\right]\ge p \qquad  \forall t\in\set T\, .
\end{align}
\end{lem}

\noindent\Proof  ~Let $B_{\delta, \vec t}   :  =  \{  \hat {\vec t} \in \set T  ~|~    \|  \hat{\vec t}  -  \vec t  \|  \le \delta\}$ the Euclidean ball of radius $\delta$ centred at $\vec t$, and let $\chi_{\delta,\vec t}$ the characteristic function of $B_{\delta,\vec t}$.   
By definition of $\delta (p,\vec t,\widehat{\vec T})$, 
there exists a sequence $\{\delta_k\}_{k\in \N}$ such that $\lim_{k\to \infty}\delta_k  =  \delta (p,\vec t,\widehat{\vec T})$, and 
\begin{align}\label{bbbb}
\int  p(\d \hat {\vec t}|\vec t)  \,  \chi_{\delta_k,  \vec t}  (\hat{\vec t})  \,  =   \mathbf{Pr}\left[  \|\widehat{\vec T}-\vec t \|\le \delta_k  \right]\ge p   \qquad
\forall k\in \N
\end{align} 
Note that the sequence  $\{\delta_k\}_{k\in \N}$ can be chosen without loss of generality to be monotonically decreasing, that is, satisfying the condition  $\delta_{k+1} \le  \delta_k$ for every $k\in\N$.
Since the sequence $\{\chi_{\delta_k,\vec t}\}_{k\in \N}$ 
converges pointwise to $\chi_{\delta (p,\vec t,\widehat{\vec T}),\vec t}$, 
and is dominated by the integrable function $\chi_{\delta_1,  \vec t}$, the dominated convergence theorem implies 
\begin{align}
 \lim_{k\to \infty}  \,  \int  p(\d \hat {\vec t}|\vec t)  \,  \chi_{\delta_k,  \vec t}  (\hat{\vec t})      
= & \int  p(\d \hat {\vec t}|\vec t)  \,  \chi_{\delta  (p,\vec t,\widehat{\vec T})  \vec t}  (\hat{\vec t})    
\nonumber \\
=  &\mathbf{Pr}\left[  \|\widehat{\vec T}-\vec t \|\le \delta  (p,\vec t,\widehat{\vec T})  \right]
\label{cccc}
\end{align}
Combining Eqs. (\ref{bbbb}) and (\ref{cccc}), we obtain  
\begin{align}\label{bbbb}
\mathbf{Pr}\left[  \|\widehat{\vec T}-\vec t \|\le \delta (p,\vec t,\widehat{\vec T}) \right]
\ge p   \, .
\end{align} 
Finally, recall that, by definition, one has  $\delta_p \ge  \delta (p,\vec t,\widehat{\vec T})$ 
for every $\vec t \in\set T$.  Hence, we have 
\begin{align}
\mathbf{Pr}\left[  \|\widehat{\vec T}-\vec t \|\le \delta_p \right]  \ge \mathbf{Pr}\left[  \|\widehat{\vec T}-\vec t \|\le \delta (p,\vec t,\widehat{\vec T}) \right]  \ge p \, .
\end{align}
\qed

\medskip 

We are now ready to provide the proof of Lemma \ref{lemma-distortion}.  

\subsection{Proof of Lemma \ref{lemma-distortion}}
Consider an \red{estimate} $\widehat{\vec T}$ for a $v$-dimensional random variable $\vec T$.  The mutual information between the two variables satisfies the bound 
\begin{align}
I(\widehat{\vec T}:\vec T)&=H(\vec T)-H(\vec T|\widehat{\vec T})
=H(\vec T)-H(\vec T-\widehat{\vec T}|\widehat{\vec T})\nonumber\\
&\ge H(\vec T)-H(\vec T-\widehat{\vec T})\label{app-entropy},
\end{align}
where $H(\vec X)$ and $H(\vec X|\vec Y )$ denote the differential entropy and the conditional differential entropy of two generic random variables $\vec X$ and $\vec Y$, respectively, and the inequality holds because conditioning of classical random variables does not increase the entropy. 

An upper bound on the differential entropy $H({\vec T}  -  \widehat{\vec T})$  can be obtained by maximizing it over the probability distributions  for the random variable $\vec T  -  \widehat{\vec T}$, under  the constraint   
\begin{align}\label{basta}
\mathbf{Pr}\left[  \|\widehat{\vec T}-\vec T \|\le \delta_p\right]\ge p.
\end{align}
Recall that the domain of the random variable $\vec T$ has the form $\set{T}=\prod_{i=1}^v    [ t_{0,i},  t_{1,i} ]$, and so does $\hat{\vec T}$. Therefore,  the domain of the random variable $\vec T-\widehat{\vec T}$ is always contained in the set $\set{T}'=\prod_{i=1}^v[t_{0,i}-t_{1,i},t_{1,i}-t_{0,i}]$, whose size is $|\set{T}'|=2^v|\set{T}|$. 

Now, let us consider maximizing the differential entropy $H(\vec X)$ under the constraint
\begin{align}
\mathbf{Pr}\left[  \|\vec X \|\le \delta_p\right]\ge p
\end{align}
over any random variable $\vec{X}$ on $\set{T}'$. 
We choose an additional parameter $s\ge 0$ as the following way; 
\begin{align}\label{H613}
\mathbf{Pr}\left[  \|\vec X \|\le \delta_p\right]= p+s.
\end{align}
Since we are potentially considering a broader class of distributions, the maximum of $H(\vec X)$ is an upper bound on $H({\vec T}-\widehat{\vec T})$.
Using Lagrange multipliers, one can show that $H(\vec X)$ is maximized when $\vec X$ has a piecewise-constant probability density function.
Differentiating   the Lagrangian $\spc{L}=-\int p(\vec x)\log p(\vec x)+\lambda_0(\int p(\vec x)-1)+\lambda_1(\int_{|\vec x|\le\delta} p(\vec x)-p-s)$ 
shows that $p(\vec x)=2^{\lambda_0+\lambda_1-1/\ln 2}$ for $|\vec x|\le \delta$ and $p(\vec x)=2^{\lambda_0-1/\ln 2}$ for $|\vec x|> \delta$. $H(\vec X)$ is then maximized by distributions of the piecewise-constant form:
\begin{align}\label{pdist}
p(\vec x)=\left\{\begin{matrix}\frac{p+s}{B_{v,\delta_p}}&\quad &|\vec x|\le\delta_p\\ 
\\
\frac{1-p-s}{2^v|\set{T}|-B_{v,\delta_p}}& &|\vec x|>\delta_p,
\end{matrix}\right.
\end{align}
where   $B_{v,\delta_p}=\frac{(\sqrt{\pi}\delta_p)^v}{\Gamma(v/2+1)}$ denotes the volume of a $v$-ball with radius $\delta_p$.  
Hence, the condition \eqref{H613} implies the inequality;
\begin{align}
H(\vec X)\le 
f(s):=&
(p+s)\log\left(\frac{B_{v,\delta_p}}{p+s}\right)
\nonumber \\
&+(1-p-s)\log\left(\frac{2^v|\set{T}|-B_{v,\delta_p}}{1-p-s}\right) \, .
\end{align}
Since the condition \eqref{H614} implies 
$\log\left(\frac{1-p-s }{p+s}\right)
\le\log\left(\frac{1-p }{p}\right)
\le \log\left(\frac{2^v|\set{T}|-B_{v,\delta_p}}{B_{v,\delta_p}}\right) 
$, we have
\begin{align}
&\frac{d f(s)}{ds}
=
\log\left(\frac{B_{v,\delta_p}}{p+s}\right)
-\log\left(\frac{2^v|\set{T}|-B_{v,\delta_p}}{1-p-s}\right) 
\nonumber \\
=&
\log\left(\frac{1-p-s }{p+s}\right)
-\log\left(\frac{2^v|\set{T}|-B_{v,\delta_p}}{B_{v,\delta_p}}\right) 
\ge 0.
\end{align}
Thus, the maximum of the RHS is achieved when $s=0$.
Therefore, due to the condition \eqref{basta}, $H(\vec T-\widehat{\vec T})$ can be bounded as
\begin{align}\label{app-h}
H(\vec T-\widehat{\vec T})\le p\log\left(\frac{B_{v,\delta_p}}{p}\right)+(1-p)\log\left(\frac{2^v|\set{T}|-B_{v,\delta_p}}{1-p}\right) \, .
\end{align}
Combining Eq. (\ref{app-entropy}) and Eq. (\ref{app-h}) we get 
\begin{align}\label{bis}
I(\widehat{\vec T}:\vec T) \ge & H(\vec T)   -  p\log\left(\frac{B_{v,\delta_p}}{p}\right)\nonumber \\
&-(1-p)\log\left(\frac{2^v|\set{T}|-B_{v,\delta_p}}{1-p}\right)  \, ,
\end{align}
which coincides with Eq. (\ref{mutualinfo1}) in the main text. 

Furthermore,  substituting the expression   $B_{v,\delta_p}=\frac{(\sqrt{\pi}\delta_p)^v}{\Gamma(v/2+1)}$  into Equation (\ref{bis}), we get:
\begin{align}
I(\hat{\vec T}:\vec{T}) 
\ge & -pv\log(\sqrt{\pi}\delta_p)+H(\vec T)+p\log\Gamma\left(\frac v2+1\right) \nonumber\\
& -h(p)-(1-p)\log(2^v|\set{T}|-B_{v,\delta_p})\nonumber\\
\ge &-pv\log(\sqrt{\pi}\delta_p)+H(\vec T)+p\log\Gamma\left(\frac v2+1\right)  \nonumber\\
&-h(p) -(1-p)\log(2^v|\set{T}|) \,,
\end{align}
where $h(p)= -p\log p  -  (1-p) \log (1-p)$ denotes the binary entropy. The second inequality comes from the monotonicity of logarithm.

\section{Proof of Lemma \ref{lem:infinitesimal}}\label{app:2Renyi} 
\subsection{2-R\'{e}nyi divergence for quantum channels}
The proof of Lemma \ref{lem:infinitesimal} uses a few properties of the 2-R\'{e}nyi divergence for quantum channels, reviewed in the following.

First, the  2-R\'{e}nyi divergence of two quantum channels  $\map A$ and $\map B$ has a finite value if and only if the following condition holds: 
\begin{cond}\label{cond:support} The support of $\Choi ({\map{B}})$ contains the support of $\Choi ({\map{A}})$. 
\end{cond}

When Condition \ref{cond:support} is satisfied, the  2-R\'{e}nyi divergence has an explicit expression provided by the following lemma:  
\begin{lem}\label{lem-form}
Let $\map A$ and $\map B$ be two quantum channels satisfying Condition \ref{cond:support}, and let $A:  =   \Choi  ({\map{A}}) $ and $B:  =   \Choi  ({\map{B}})$ be their Choi operators. Then, one has 
  \begin{align}\label{RLD}
{D_2(\map{A}||\map{B})}=  \log \left\|\Tr_{\spc{H}_{\rm out}} \,  \left[ A B^{-1} A \right] \right\|_{\infty},
\end{align}
where $   \|  X  \|_\infty   :  =  \sup_{|\psi\>  \in \spc H,  |\psi\> \not = 0}      \,  \|   X  |\psi\>  \|  /\|  |\psi\>  \|$ denotes  the operator norm of a generic operator $X  \in L  (\spc H)$. \end{lem}

\noindent\Proof     ~Lemma \ref{lem-form} is  as a special case of 
\cite[Item 2 of Theorem 3]{fang2019geometric} with $\alpha=2$. For the reader's convenience, we provide here a self-contained proof using only elementary techniques.  

 Let $|\psi\>   =  \sum_{m=1}^r  \,  \sqrt {p_m} \,  |\alpha_m\>  \otimes |\beta_m\>$ be a Schmidt representation of the state $|\psi\>$ in Equation (\ref{def-D2}).  For  $\map C\in  \{\map A, \map B\}$, one has  $(\map C \otimes \map I)   (\psi)   =       (I_{\rm out}  \otimes   F )  \Choi (\map C) (I_{\rm out}  \otimes   F )^\dag$,  with $F:  =\sum_m  \,  \sqrt {p_m}  \,  |\beta_m\>\<\overline \alpha_m|$.     When  $F$ is invertible,      the $2$-R\'{e}nyi divergence between the output states is well-defined thanks to Condition \ref{cond:support}, and  can be written as 
\begin{align}
& D_2\Big( \,  (\map{A}\otimes\map{I}) \,(\psi)  \Big \| \,  (\map{B}\otimes\map{I} )\, (\psi)\Big)    \nonumber\\
 =& \log \Bigg\{  \Tr \Bigg[     \Big(    (I_{\rm out}  \otimes   F )  A  (I_{\rm out}  \otimes   F )^\dag \Big)^2    
\nonumber\\
 &\hspace{18ex} \cdot\Big(   (I_{\rm out}  \otimes   F )  B (I_{\rm out}  \otimes   F )^\dag \Big)^{-1}  \Bigg]\Bigg\}    \nonumber \\
=& \log \left\{  \Tr \left[    A  (I_{\rm out}  \otimes   F^\dag F )   A  B^{-1}     \right]  \right\}\nonumber 
\\
=& \log \left\{   \Tr \left[    (I_{\rm out}  \otimes   F^\dag F )   A B^{-1} A    \right] \right\}\nonumber \\
 =&  \log \left\{  \Tr    \left[     F^\dag F  \,  \Tr_{\rm out}  [ A B^{-1} A]  \,     \right]\right\}   \, .  \label{finale} 
\end{align}
When $F$ is not invertible, one can represent $F$ as a limit of a sequence of invertible operators, and Equation (\ref{finale}) still holds by continuity.  

Note that, for a generic pure state  $|\psi\>=  \sum_m \,\sqrt{p_m}  \,  |\alpha_m\>  |\beta_m\> $, the operator $F^\dag F   =  \sum_m  \,  p_m \,  |\overline \alpha_m\>\<\overline \alpha_m|$ is a generic mixed state.   Hence, one has 
 \begin{align}
& D_2(\map{A}||\map{B}) \nonumber\\
=& 
\sup_{\psi\in\operatorname{PurSt}(\spc{H}_{\rm in}\otimes \spc{H}_{\rm in})} \quad   D_2\Big( \,  (\map{A}\otimes\map{I}) \,(\psi)  \Big \| \,  (\map{B}\otimes\map{I} )\, (\psi)\Big)  \nonumber\\
\nonumber =& \sup_{\rho  \in \operatorname{St} (\spc H_{\rm in}) }  \,    \log \left\{   \Tr    \left[     \rho   \,  \Tr_{\rm out}  [  A B^{-1} A   \,     \right]\right\} 
=  \log    \left\|   \,   \Tr_{\rm out}  [  A B^{-1} A ]  \,  \right\|_\infty \, .
\end{align}
\qed

\begin{cor}\label{cor:additivity}
If channels $\map{A}_i$ and $\map{B}_i$ satisfy Condition  \ref{cond:support} for $i  \in  \{1,2\}$, then the $2$-R\'{e}nyi divergence  has the additivity property   
\begin{align}\label{additivity}
D_2\left(\map{A}_1\otimes\map{B}_1||\map{A}_2\otimes\map{B}_2\right)=D_2\left(\map{A}_1||\map{A}_2\right)+D_2\left(\map{B}_1||\map{B}_2\right) \, .
\end{align}
\end{cor}
\Proof ~Immediate  from Equation (\ref{RLD}).  Corollary \ref{cor:additivity} is  a special case of \cite[Item 3 of Theorem 3]{fang2019geometric} with $\alpha=2$. It can also  be derived in a similar way as  Theorem 2 of \cite{hayashi2011comparison}. 
  \qed

\begin{lem}\label{lem:posi}
Let $\map A$ and $\map B$ be two quantum channels satisfying Condition \ref{cond:support}, and let $A:  =   \Choi  ({\map{A}}) $ and $B:  =   \Choi  ({\map{B}})$ be their Choi operators. Then, one has
\begin{align}\label{posi}
\Tr_{{\rm out}}  \left[ A B^{-1} A  \right]    \ge I_{\rm in}
\end{align}
and
\begin{align}
\label{RLD2}
&2^{D_2(\map{A}||\map{B})}-1 
=\left\|\Tr_{\spc{H}_{\rm out}} \, \left[   (A- B)  \, B^{-1}  (A-B)\right]  \, \right\|_{\infty}
 \, .
\end{align}
\end{lem}

\Proof ~For a generic unit vector  $|\alpha\>  \in  \spc H_{\rm in}$,  one has 
\begin{align}
\nonumber & \<\alpha|  \, \Tr_{{\rm out}}  \left[ A B^{-1} A  \right]    \, |\alpha \>    =   \Tr \left [  |\alpha\>\<\alpha|\,  \Tr_{\rm out} \,\left[  A B^{-1} A   \right]   \right]  \\
   =&  2^{D_2\Big( \,  (\map{A}\otimes\map{I}) \,(\psi)  \Big \| \,  (\map{B}\otimes\map{I} )\, (\psi)\Big) }   
 \ge  1   \, ,\label{bbb}
\end{align}
where $|\psi\>  := |\overline \alpha\>  |\alpha\>  $ and the second equality follows from Equation (\ref{finale}) with $F  = |\alpha\>\<\alpha| $, while the inequality follows from the fact that the   2-R\'{e}nyi divergence  of quantum states is non-negative. 
  Since the vector $|\alpha\>$ is generic, Equation (\ref{bbb}) proves Equation  (\ref{posi}). 
 
 To prove Equation  (\ref{RLD2}), consider first the case where $\Choi (\map A)$ and $\Choi (\map B)$ are invertible.       In this case, one has   
 \begin{align} 
\nonumber  \Tr_{\rm out}  \left[     (A-  B) \,  B^{-1} \,  (A-B)\right]    &  =   \Tr_{\rm out}  \left[    A B^{-1} A   +  B  -  2 A\right]  \\
&  =     \Tr_{\rm out}  \left[  A B^{-1} A     \right]  -  I_{\rm in}  \, ,\label{ccc}
\end{align}
  having used the relation  $\Tr_{\rm out}  [A ]   = \Tr_{\rm out}  [B ] =  I_{\rm in} $.    

Hence, one has
\begin{align}
& \nonumber \Bigg\|\Tr_{\spc{H}_{\rm out}} \, \Bigg[ \big(\Choi  ({\map{A}})-\Choi (\map{B})  \big)
\Choi^{-1}  ({\map{B}})  \\
&\hspace{20ex}\cdot  \big(\Choi  ({\map{A}})-\Choi (\map{B})  \big)\Bigg]  \, \Bigg\|_{\infty}  \nonumber \\
 =&\left \|  \Tr_{\rm out}  \left[    A B^{-1} A   \right]  -  I_{\rm in}  \right\|_{\infty}\nonumber\\
   =&  \left \|  \Tr_{\rm out}  \left[   AB^{-1} A  \right]   \right \|_{\infty}- 1
   = 2^{D_2(\map{A}||\map{B})}-1 \, ,
\end{align}
the second equality following from Equation (\ref{posi}), and the third equation following from Equation (\ref{RLD}). \qed

\medskip 


\subsection{Proof of Lemma \ref{lem:infinitesimal}}  
Since the support of the Choi operator of  the channels $\{\map C_{\vec t}\}$ is independent of $\vec t$, we can apply Lemma \ref{lem:posi} with   $\map{A}=\map{C}_{\vec{t}'}$ and  $\map{B}=\map{C}_{\vec{t}}$, obtaining 
\begin{align*}
2^{D_2(\map{C}_{\vec{t}'}||\map{C}_{\vec{t}})}-1
&=\left\|\Tr_{{\rm out}}  [  \,(  C_{\vec t'}  -  C_{\vec t}  )  \,  C_{\vec t}^{-1} \, (  C_{\vec t'}  -  C_{\vec t}  )\, ]  \,\right\|_\infty \, ,
\end{align*}
where $C_{\vec t}$ and $C_{\vec t'}$ are the Choi operators of channels $\map C_{\vec t}$ and $\map C_{\vec t'}$, respectively. 

Defining  $\epsilon :  =  \|  \vec t'  -  \vec t\|$ and $  \vec s  :  =    (  \vec t'  - \vec t)/ \|  \vec t'  -  \vec t\|$,  the Taylor expansion of the operator $C_{\vec t'}$ 
\begin{align}
& 2^{D_2(\map{C}_{\vec{t}'}  ||\map{C}_{\vec{t}})}-1
\nonumber \\
=& \Big\|\Tr_{\spc{H}_{\rm out}}
\Big(\epsilon \,  \sum_i\frac{\partial  C_{\vec t}}{\partial t_i}  \, s_i
+O(\epsilon^2)\Big)
  C_{\vec t}^{-1}\, 
\Big(   \epsilon  \,  \sum_j\frac{\partial C_{\vec t} }{\partial t_j} \, s_j
+O(\epsilon^2)\Big)\Big\|_{\infty}\nonumber\\
=& \epsilon^2 \left\|\sum_{i,j}s_i s_j\Tr_{\spc{H}_{\rm out}}\left(\frac{\partial   C_{\vec t} }{\partial t_i}  C_{\vec t}^{-1}  \,   \frac{\partial  C_{\vec t} }{\partial t_j}\right)\right\|_{\infty}
+O\left(\epsilon^3\right)  \, ,\label{appendix-inter1}
\end{align}
where the $O\left(\epsilon^3\right)$ can be uniformly bounded with respect to $\vec s$. 

Substituting Eq. \eqref{RLD1} into Eq. \eqref{appendix-inter1}, one gets
\begin{align}
2^{D_2 (\map{C}_{\vec{t}+\epsilon \vec{s}}|| \map{C}_{\vec{t}} )}
-1
\le
\epsilon^2
J^{\rm R}_{\map{C}_{\vec{t}}}
+O\left(\epsilon^3\right) \, ,
\label{eq-inter1}
\end{align}
which proves the desired result.  \qed

\bibliographystyle{IEEEtran}
\bibliography{ref}

\end{document}